\documentclass{ejm}%
\usepackage{amsmath}
\usepackage{amsfonts}
\usepackage{amssymb}
\usepackage{graphicx}%
\newtheorem{theorem}{Theorem}

\newtheorem{lemma}[theorem]{Lemma}

\newtheorem{oproblem}[theorem]{Open problem}
\newtheorem{proposition}[theorem]{Proposition}

\title[Localized patterns in the Brusselator]
{Mesa-type patterns in the one-dimensional Brusselator and their stability}

\author[Theodore Kolokolnikov, Thomas Erneux and Juncheng Wei]{%
  T.\ns K\ls O\ls L\ls O\ls K\ls O\ls L\ls N\ls I\ls K\ls O\ls V$~^\dag$, \ns
  T.\ns E\ls R\ls N\ls E\ls U\ls X$~^\dag$ \ns
  \and  J.\ns W\ls E\ls I$~^\ddag$
}

\affiliation{$~^\dag$\ Universit\'e Libre de Bruxelles, Optique Nonlin\'eaire
Th\'eorique, Campus Plaine, C.P. 231, 1050 Brussels, Belgium\\
tkolokol\symbol{64}gmail.com, terneux\symbol{64}ulb.ac.be

$~^\ddag$\ Department of Mathematics,
  The Chinese University of Hong Kong,
  Shatin, Hong Kong \\
wei\symbol{64}math.cuhk.edu.hk}
\pubyear{2005}
\volume{000}
\pagerange{\pageref{firstpage}--\pageref{lastpage}}

\begin{document}

\maketitle

\begin{abstract}
The Brusselator is a generic reaction-diffusion model for a tri-molecular
chemical reaction. We consider the case when the input and output reactions
are slow. In this limit, we show the existence of $K$-periodic, spatially
bi-stable structures, \emph{mesas}, and study their stability. Using singular
perturbation techniques, we find a threshold for the stability of $K$ mesas.
This threshold occurs in the regime where the exponentially small tails of the
localized structures start to interact. By comparing our results with Turing
analysis, we show that in the generic case, a Turing instability is followed
by a slow coarsening process whereby logarithmically many mesas are
annihilated before the system reaches a steady equilibrium state. We also
study a ``breather''-type instability of a mesa, which occurs due to a Hopf
bifurcation. Full numerical simulations are shown to confirm the analytical results.

\end{abstract}

\label{firstpage}

\section{Introduction}

In 1952, Turing proposed that the formation of spatial patterns during
morphogenesis could be explained in terms of the instability of a homogeneous
steady-state solution of a reaction-diffusion network describing the evolution
of a set of morphogens \cite{Turing}. Turing himself illustrated his ideas on
two chemical models. Turing's original work is primarily concerned with the
stability analysis of the homogeneous steady-state solution of the rate
equations for the interacting morphogens \cite{NP}. The main point of
biological interest, however, is whether stable spatial structures may be
generated beyond the instability, i.e., whether the rate equations admit
stable (and positive) inhomogeneous solutions exhibiting the most
characteristic features of morphogenetic patterns.

This point has been taken up seriously in the early seventies. More systematic
numerical studies of Turing's model have been performed showing irregular
spatial structures \cite{bard}. This led to serious reservations about the
relevance of Turing's theory in developmental biology, particularly its
ability to generate regular patterns. But these criticisms originate from a
somewhat unfortunate choice of Turing's example and they do not touch the
essential points of Turing's theory. More specifically, the obvious
requirement that the rate equations must admit positive and bounded solutions
is not satisfied in Turing's example \cite{erneux2}.\ 

Because of this misunderstanding on Turing's fundamental contribution, other
two-variable models satisfying the law of mass action have been explored. In
1968, Prigogine and Lefever \cite{PL} introduced a two variable system that
exhibits an autocatalytic reaction (called the ``Brusselator''). The
simplicity of the rate equations motivated analytical and numerical studies
which showed the existence of stable structures \cite{auchmuty}. The
Brusselator is based on the following intermediate reactions for the two
chemical intermediates $X$ and $Y$:
\begin{equation}
A \rightarrow X, ~~~ C + X \rightarrow Y + F, ~~~ 2X + Y \rightarrow3 X, ~~~ X
\rightarrow E.
\end{equation}

\begin{figure}[tb]
\begin{center}
\setlength{\unitlength}{1in} \begin{picture}(6.5,3.0)(0.8,0)
\put(0.16,0.25){\includegraphics[width=3.5in]{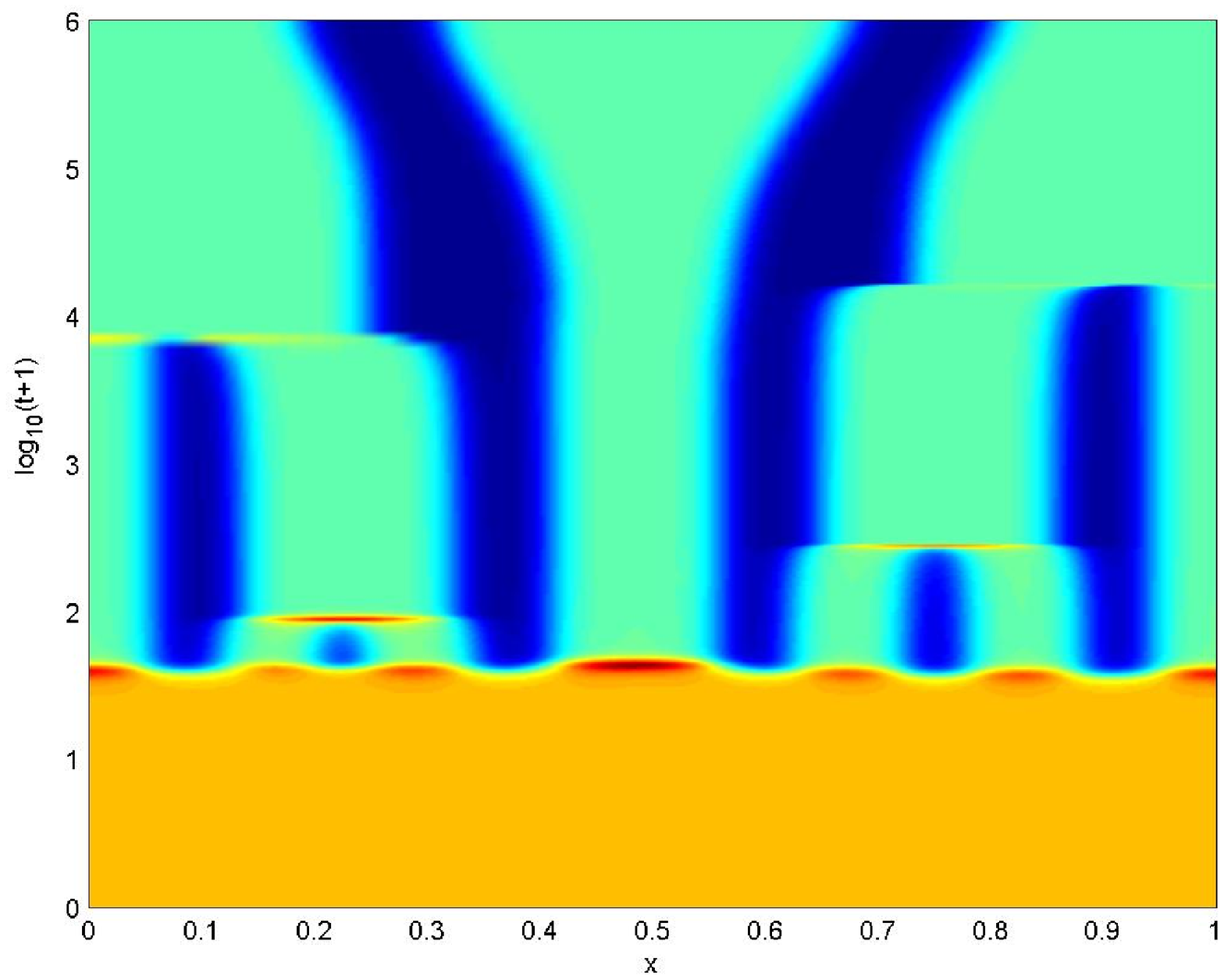}}
\put(3.333,0.25){\includegraphics[width=3.5in]{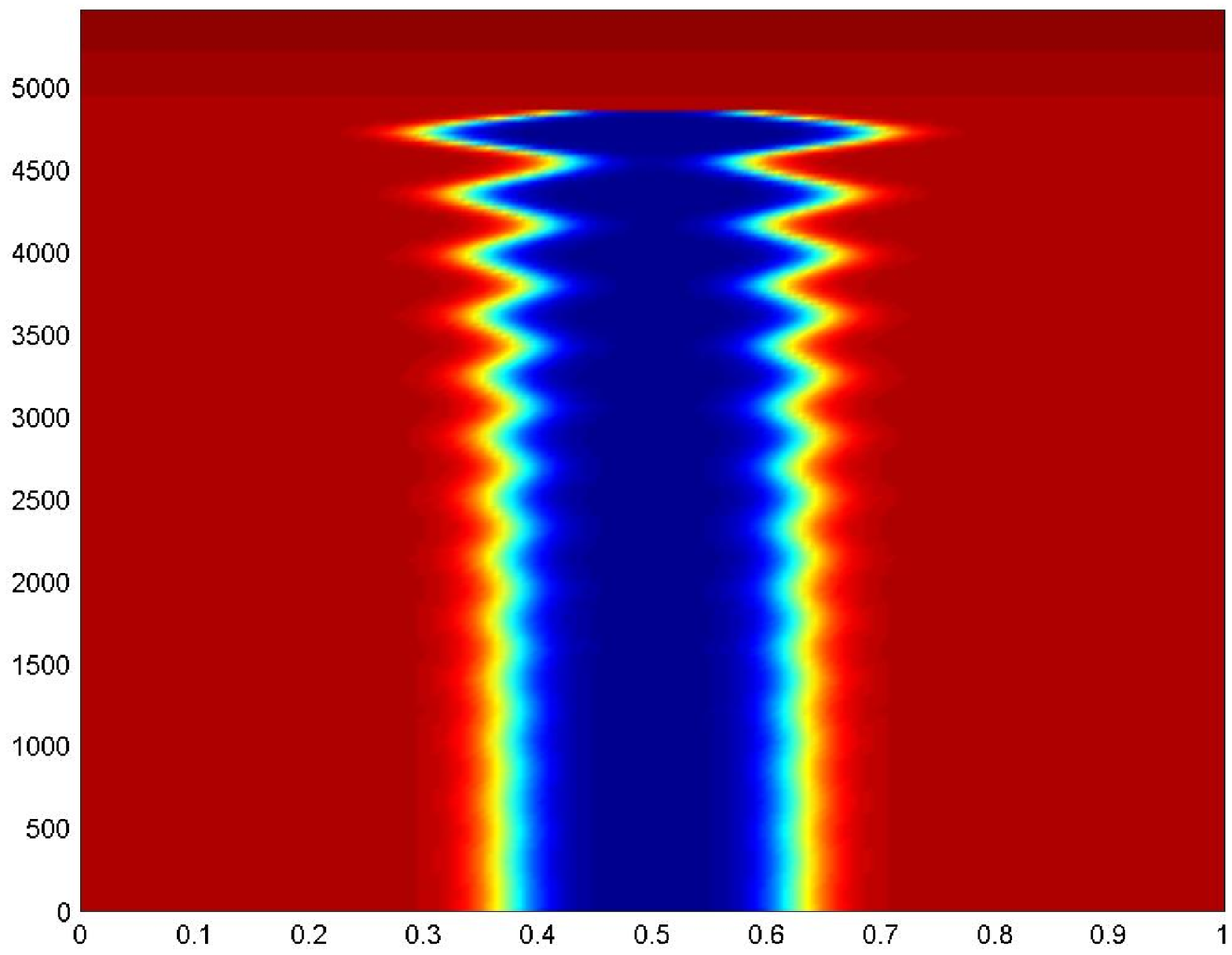}}
\put(1.7,0.1){(a)}
\put(5.0,0.1){(b)}
\end{picture}
\end{center}
\caption{(a) Turing instability and mesa-type localized structures. The
parameters are $A=1,B=8,~\varepsilon=10^{-4},~D=10,~\tau=10.$ Initial
condition was set to $u=A, v=A/B$, perturbed by a very small random noise.
Note the logarithmic scale for time. Initial Turing instability triggers a
$k=7$ mode at time $t \sim50$. Thereafter a coarsening process takes place
until there are only two mesas left. (b) Slow-time oscillatory instability of
a single spike solution to (\ref{6dec4:04}). The parameter values are
$A=1,~B=8,~D=10,~\varepsilon=0.00025,~\tau=0.999.$}%
\label{fig:turing}%
\end{figure}

The global reaction is $A+C\rightarrow F+E$ and corresponds to the
transformation of inputs products $A$ and $C$ into output products $F$ and
$E.$ We assume (without loss of generality) that the rate constants for the
first and last step are equal to $r$ whereas the intermediate rate constants
are one. The rates equations then become
\begin{align}
X_{t}  &  =D_{x}X_{xx}+r A-C X+X^{2}Y-rX\\
Y_{t}  &  =D_{y}Y_{xx}+CX-X^{2}Y.
\end{align}

In this paper we concentrate on the case where the input and the output reaction
steps are slow in comparison to the intermediate steps, so that $r$ is small. In the
absence of diffusion terms, this implies that the total concentration $X+Y$
is a slowly changing quantity. Linear stability reveals the presence of a periodic
orbit which arise through a Hopf bifurcation of the steady state $X=A, Y=C/A$. In
the limit $r \rightarrow0$ such a periodic orbit consists of a slow and a fast
phase, so that a pulse train can be observed.

There is a large body of literature on the space-independent Brusselator
($D_{x}=0=D_{y}$) and its extensions. Under different assumptions on the
parameters, the equilibrium state is only marginally stable and an addition of
small random perturbations (or small periodic forcing) can lead regular to
large-amplitude pulses \cite{OP}, \cite{MVE}. Other authors have looked at
coupled brusselator systems which can exhibit chaos and syncrhonization
\cite{Tyson}, \cite{YuGumel}, \cite{ChakravartiMarekRay}.

Since the discovery of spatial patterns in 1970's, various Turing patterns in
the Brusselator were studied both numerically and analytically in one, two and
three space dimensions. These include spots, stripes, labyrinths and and
hexagonal patterns \cite{Erneux}, \cite{NP}, \cite{deWit}, \cite{Pena},
\cite{deWit3D}, oscillatory instabilities and spatio-temporal chaos
\cite{deWitChaos}, \cite{YZE}. While Turing analysis and its weakly nonlinear
extensions have been successful at detecting and classifying possible pattern
types, its range of applicability is limited. Indeed Turing patterns are
assumed to be small sinusoidal perturbations of a homogeneous state. In practice
however, many patterns are localized and contain sharp transitions such as spikes
and kinks.  As we will see, this is particularly the case under the assumption of
small $r$. In this paper we study \emph{mesa}-type patterns.\footnote{Mesa means
table in Spanish; it is also a name given to square boulders which are found in the
Colorado desert. The use of this term was suggested by Fife \cite{Fife}.} These are
box-like patterns that join two flat regions of space with sharp transition layers,
such as shown in Figure \ref{fig:1}. Such patterns are not amenable to Turing
analysis since they are far away from the homogeneous state.

Previous mathematical efforts on the formation of localized patterns concentrated on
two variable reaction-diffusion models where the dynamics is controlled by the
interaction between a slow and a fast variable.  For the Brusselator without
diffusion terms and with $r$ small, the total concentration $X+Y$ plays the role of
a slow variable. As a result, localized patterns patterns are expected when
diffusion is permitted.

For our analysis it will be convenient to use the following scaling,
\begin{equation}\label{6dec4:04}
\begin{aligned} \tau u_{t} & =D \varepsilon u_{xx}+\varepsilon A-Bu+u^{2}v-\varepsilon u\\ v_{t} & =D \varepsilon v_{xx}+Bu-u^{2}v, \end{aligned}
\end{equation}
where
\begin{equation}
\tau=\frac{D_{y}}{D_{x}},~~~\varepsilon=r\frac{D_{y}}{D_{x}},~~~B=C\frac
{D_{y}}{D_{x}},~~~D=\frac{D_{x}}{r},
\end{equation}
\begin{equation}
u=X,~~~~v=\tau Y
\end{equation}
and the spatial domain is $x\in\lbrack0,1]$ with the zero flux boundary
conditions
\begin{equation}
u_{x}=0=v_{x}~~\mbox{at}~x=0~\mbox{and}~1.
\end{equation}

In this work we assume the following conditions,
\begin{align}
r \ll D_{x} \le O(D_{y}) \ll1,~~~~ A=O(1), ~~ C = O\left(  \frac{D_{x}}{D_{y}%
}\right)  .
\end{align}
In terms of our scaling we have
\begin{equation}
\label{epsD}\varepsilon D\ll1,~~~D\gg1 ; ~~~~~O(A)=1=O(B),
\end{equation}
\begin{equation}
O(\tau) \ge1.
\end{equation}

As a motivation, let us present two numerical examples. On Figure
\ref{fig:turing}.a we show a typical time evolution over a very long time
interval. The homogeneous steady state is unstable because of Turing
instability but the expected Turing sinusoidal pattern only appears after a
delay ($t > 50$). Turing's structure then gradually deteriorates and
relatively quickly moves into a new pattern formed by several localized
mesa-type structures. They then undergo a coarsening process over a
logarthmically long time-scale and, eventually, only two mesas remain.
Turing's analysis can be used to predict the first pattern ($t \sim50$) but it
cannot anticipate the coarsening process or the number of final mesas. As we
shall demonstrate, in the case $B > A^{2}$, the Turing pattern is
characterized by a wave number proportional to $k=O\left(  \frac{1}%
{\sqrt{D\varepsilon}}\right)  $ while the number of stable long-time mesas has
the order $K=O\left(  \frac{1}{\sqrt{D\varepsilon} \ln\frac{1}{\varepsilon}%
}\right)  $, so that $k$ is logarithmically bigger than $K$. This is the
underlying cause of the coarsening process observed in Figure \ref{fig:turing}.a.

In the second experiment shown in Figure \ref{fig:turing}.b, a single mesa
undergoes a ``breather''-type oscillatory instability which eventually leads
to its extinction. Both coarsening and the breather instability occur at a
slow timescale. The main goal of this work is to describe these instabilities analytically.

Our results are related to the study of bistable systems, see for example
\cite{RW}, \cite{GMP}, \cite{N0} \cite{N1}, \cite{N2}, \cite{N3}, \cite{N4},
\cite{DI}. Mesa patterns also appear in the FitzHugh-Nagumo model \cite{GMP},
certain phase separation models such as Cahn-Hilliard, Allen-Cahn, \cite{ACF},
\cite{AFS}, \cite{C} and block-copolymers \cite{RW}. For these systems the
resulting spectral problem has small eigenvalues, also called critical
spectra, that tend to zero with the thickness of the interfaces. Typically $k$
such layers are stable \cite{N3}. However as we show in this paper, if the
number of layers is excessively large, instabilities can occur. This happens
when the exponentially small interaction outside the interface locations
cannot be ignored. Our main new contribution is to study this interaction, and
to show that it has a destabilizing effect.

\subsection{Summary of main results}

We now summarize our main results. We first describe the shape of the
equilibrium $K$ mesa solutions. In Section \ref{sec:2} we show the following.

\begin{figure}[tb]
\begin{center}
\setlength{\unitlength}{1in} \begin{picture}(5.0,3.1)(0,0)
\put(2.5, 0){$x$}
\put(0, 1.6){$v$}
\put(0.2,0.2){\includegraphics[width=4.5in]{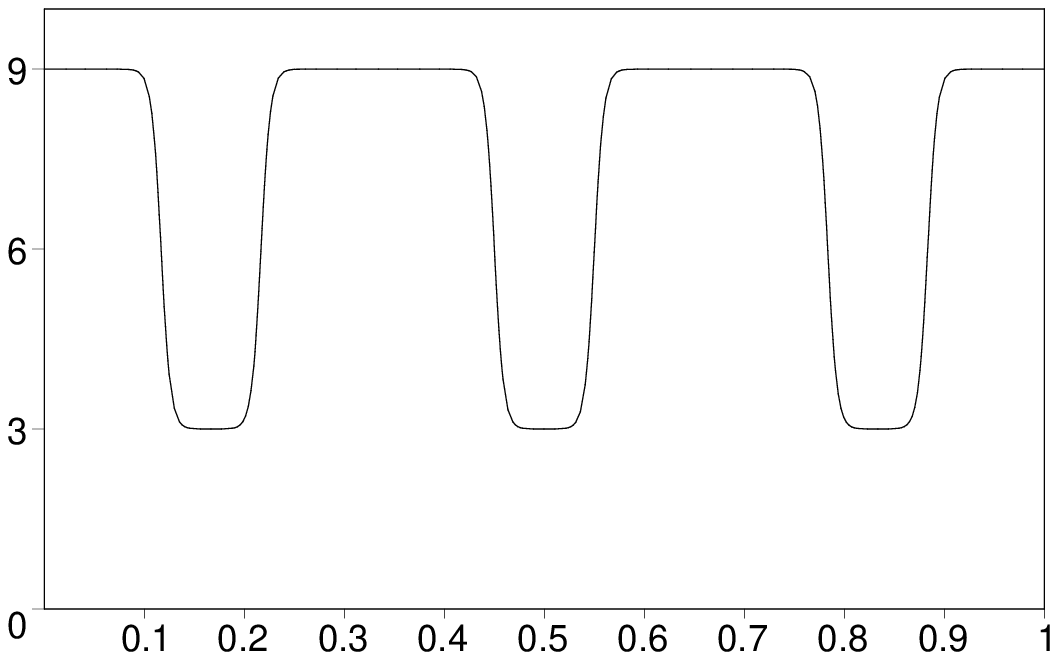}}
\end{picture}
\end{center}
\caption{An example of a three-mesa equilibrium solution for $v$. Here, $K=3,~
A=2,~B=18,~\varepsilon D=0.02^{2},$ $w_{0}=9$, $l=0.11.$ }%
\label{fig:1}%
\end{figure}

\begin{proposition}
\label{prop:1}Consider the equilibrium-state problem,%
\begin{gather}
0=\varepsilon Dv_{xx}+Bu-u^{2}v,\ \ \ \ \ \ 0=\varepsilon Du_{xx}+\varepsilon
A+u^{2}v-\left(  B+\varepsilon\right)  u,\ \ \ \ x\in\left[  0,1\right]  ,\\
u^{\prime}=0=\ v^{\prime}\text{ \ \ \ at }\ \ \ x=0\text{ and }x=1.
\end{gather}
in the limit (\ref{epsD}) and suppose that
\begin{align}
\label{15apr17:20}A^{2}<2B.
\end{align}
Then there exists a K-mesa symmetric solution to $(v,u)$ of the following form.

Let%
\begin{align}
\label{wld}w_{0}=3\sqrt{B/2},\ \ \ \ l=\frac{A}{K\sqrt{2B}},~~~~ d = \frac
{1}{K}-l,
\end{align}
\begin{equation}
\label{14apr15:50}x_{li}\equiv x_{i}-\frac{l}{2},\ \ \ x_{ri}\equiv
x_{i}+\frac{l}{2},\ \ \ x_{i}\equiv\frac{\left(  \frac{1}{2}+i\right)  }%
{K}\ \ \text{for} \ \ i=1\ldots K.
\end{equation}

For $x$ away from $x_{ri},$ $x_{li}$, we have:%
\begin{equation}
v\sim\left\{
\begin{array}
[c]{l}%
w_{0},\ \ \ \ x\in\lbrack0,1]\backslash\cup\left[  x_{li},x_{ri}\right] \\
\frac{w_{0}}{3},\ \ \ \ x\in\cup\left(  x_{li},x_{ri}\right)
\end{array}
\right.
\end{equation}
For $x$ near the interfaces $x_{ri},$ $x_{li}$ we have,%
\begin{equation}
v\sim\left\{
\begin{array}
[c]{l}%
v_{l}(x-x_{li}),\ \ \ \ x-x_{li}\leq O(\sqrt{\frac{\varepsilon D}{B}})\\
v_{r}(x-x_{ri}),\ \ \ \ x-x_{ri}\leq O(\sqrt{\frac{\varepsilon D}{B}})
\end{array}
\right.
\end{equation}
where
\begin{equation}
v_{r}=w_{0}\frac{2}{3}+w_{0}\frac{1}{3}\tanh\left(  \frac{w_{0}}{3}\frac
{x}{\sqrt{2\varepsilon D}}\right)  ,\ \ \ \ \ \ \ v_{l}=w_{0}\frac{2}{3}%
-w_{0}\frac{1}{3}\tanh\left(  \frac{w_{0}}{3}\frac{x}{\sqrt{2\varepsilon D}%
}\right)  .
\end{equation}
Finally,
\begin{equation}
u\sim w_{0}-v.
\end{equation}

\end{proposition}

A typical such solution with $K=3$ is illustrated on Figure \ref{fig:1}.

We next analyse the stability of such equilibrium. There are two distinguished
limits of interest, either $DK^{2}\ll O\left(  \frac{1}{\varepsilon{\ln
^{2}{\varepsilon}}}\right)  $ or $DK^{2}=O\left(  \frac{1}{\varepsilon{\ln
^{2}{\varepsilon}}}\right)  .$ The former is studied in Sections \ref{sec:3}
and \ref{sec:4} while the latter in Section \ref{sec:5}. In Section
\ref{sec:3} we derive rather precise results for eigenvalues, summarized in
the following theorem.

\begin{theorem}
\label{thm:largeD}Consider a $K$ mesa solution of Proposition \ref{prop:1}.
Suppose in addition that
\begin{equation}
1\ll DK^{2}\ll O\left(  \frac{1}{\varepsilon{\ln^{2}{\varepsilon}}}\right)
\text{ \ \ \ \ and \ \ }O\left(  \tau-1\right)  \gg0.
\end{equation}
Such solution is stable when $\tau-1\gg0$ and unstable when $\tau-1\ll0$.
\ There are $2K$ small eigenvalues of order $O\left(  \varepsilon\right)  ;$
all other eigenvalues are negative and have order $\geq O\left(
D\varepsilon\right)  .$ The smallest $2K$ eigenvalues are given by%
\begin{align}
\lambda_{j\pm}  &  \sim\frac{-1\pm\sqrt{1-2K^{2}dl\left[  1-\cos\left(
\frac{\pi j}{K}\right)  \right]  }}{2\left(  \tau-1\right)  }\varepsilon
\ \ \text{for}\ \ j=1\ldots K-1;\\
\lambda_{-}  &  \sim\frac{-Kl}{\tau-1}\varepsilon,\ \ \ \ \ \ \lambda
_{+}=\frac{-1}{\tau-1}\varepsilon.
\end{align}
and are all negative when $\tau>1,$ and positive when $\tau<1.$ The transition
from stability to instability occurs via a Hopf bifurcation as $\tau$ is
decreased past $\tau_{h}$ where to leading order, $\tau_{h}\sim1$.
\end{theorem}

Note that near the interfaces, the gradient changes on the order $\delta
=\sqrt{\varepsilon D}.$ In terms of $\delta,$ the scaling $DK^{2}=O(\frac
{1}{\varepsilon\ln^{2}{\varepsilon}})$ can be written as
\begin{equation}
D=O\left(  \delta^{2}\exp\left(  \frac{1}{K\delta}\right)  \right)  .
\end{equation}
Thus Theorem \ref{thm:largeD} confirms the stability of $K$ mesas when
$\tau>1$ as long as $D$ is not exponentially large in $\delta$. In the
contrary case, we derive the following result in Section \ref{sec:5}.

\begin{theorem}
\label{thm:hugeD}Suppose that
\begin{equation}
\tau>1
\end{equation}
and let
\begin{equation}
D_{K}=\frac{1}{K^{2}}D_{1}\text{ \ where \ }D_{1}\sim\left\{
\begin{array}
[c]{c}%
\frac{A^{2}}{2\varepsilon\ln^{2}\left(  \frac{12\sqrt{2}AB^{3/2}}%
{\varepsilon\left(  \sqrt{2B}-A\right)  ^{2}}\right)  },\ \ \ \ 2A^{2}<B\ \\
\frac{\left(  \sqrt{2B}-A\right)  ^{2}}{2\varepsilon\ln^{2}\left(
\frac{12\sqrt{2}}{\varepsilon A}B^{3/2}\right)  },\ \ \ \ 2A^{2}>B
\end{array}
+l.s.t.\right.
\end{equation}
Here, l.s.t. denotes logarithmically small terms. Then a $K$ mesa symmetric
equilibrium with $K\geq2$ is stable if $D<D_{K}$ and is unstable otherwise.
Moreover, a single-mesa equilibria $K=1$ is always stable. A more precise
value for $D_{1}$ is given in Proposition \ref{prop:expstab}.
\end{theorem}

\begin{figure}[ptbh]
\begin{center}
\setlength{\unitlength}{1in} \begin{picture}(5.2,3.1)(0,0)
\put(2.5, 0){$x$}
\put(2.5, 3.0){$v(x,t)$}
\put(0.1, 1.6){$t$}
\put(0.1,0.0){\includegraphics[width=5in, height=3in]{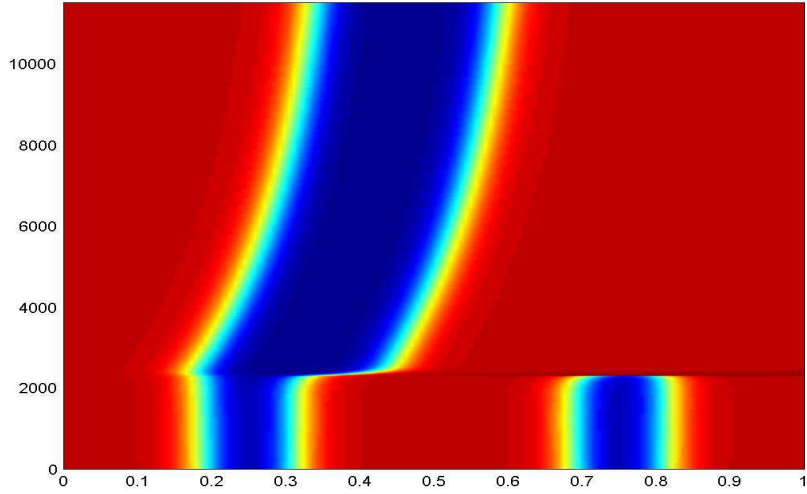}}
\end{picture}
\end{center}
\caption{ Slow-time competition instability of a two-mesa solution to
(\ref{6dec4:04}). The parameter values are $A=1,~B=8,~D=10,~\varepsilon
=0.00025,~\tau=1.3$}%
\label{fig:3}%
\end{figure}

Theorem \ref{thm:hugeD} states that the instability threshold occurs when $D$
is exponentially large. In this case the exponentially fast decay outside the
interface locations must be taken into account. Their exponentially weak
interaction is responsible for an eventual loss of stability. Figure
\ref{fig:3} illustrates this proposition. Indeed using the parameters used in
that simulation we deduce from Proposition \ref{prop:expstab} that
$D_{1}=20.96;$ so that $D_{2}=5.28.$ Since $D=10>D_{2},$ the two-mesa
equilibrium state is unstable.

Our next result is about the presence of a Hopf bifurcation when $\tau$ is
near 1.

\begin{theorem}
\label{thm:hopf} Suppose that
\begin{equation}
\text{ \ \ \ }\sqrt{\frac{B}{\varepsilon D}}\ll DK^{2}\ll O\left(  \frac
{1}{\varepsilon\ln^{2}{\varepsilon}}\right)  . \label{9mar14:03}%
\end{equation}
Let
\begin{equation}
\tau_{h_{+}}=1+\frac{1}{4D}\left(  ld-\frac{K}{3}\left(  d^{3}+l^{3}\right)
\right)  \label{9mar14:14}%
\end{equation}
where $K,d,l$ are as in Proposition \ref{prop:1}. Then a K-mesa solution
undergoes a Hopf bifurcation when $\tau=\tau_{h_{+}}.$ It is stable when
$\tau>\tau_{h_{+}}$ and unstable otherwise. When $\tau=\tau_{h_{+}},$ the
corresponding eigenvalue has value%
\begin{equation}
\lambda_{+}\sim i\sqrt{8K}\left(  \varepsilon^{3}DB\right)  ^{1/4}.
\end{equation}

\end{theorem}

Figure \ref{fig:turing}.b illustrates the type of oscillations that occur when
$\tau< \tau_{h+}$. Theorem \ref{thm:hopf} is able to predict the onset of such
oscillations even though it says nothing about whether this bifurcation is
supercritical or subcritical.

Finally we perform a Turing stability analysis of the Brusselator in Section
\ref{sec:6}. We find that in the generic case, the modes $k$ within the Turing
instability band have the order $O(\frac{1}{\delta})$ where $\delta=
\sqrt{\varepsilon D}$ is the width of the interface. On the other hand the
mesa instability threshold occurs when $K=O\left(  \frac1{\delta\ln\frac
{1}{\varepsilon}}\right)  $. It is then clear that $k \gg K$ by a
logarithmically large amount. This is the underlying reason for the coarsening
process observed in Figure \ref{fig:turing}.a.

One of the easy consequences of Turing's analysis is the existence of the
regime for which mesa-type structures are stable at the same time as the
homogeneous steady state. A more interesting question is the following.

\begin{oproblem}
Does there exist a parameter regime for which the mesa-type solution is
unstable, and in addition the homogeneous steady state $u=A,\ v=\frac{B}{A}$
is unstable with respect to the Turing instability?
\end{oproblem}

An existence of such a regime would imply spatio-temporal chaos in the
Brusselator. We answer this question in the negative for the case when
\begin{equation}
\frac{1}{\delta}\ll D\ll\delta\exp\left(  \frac{1}{\delta}\right)
,\ \ \ \ \ \ \delta=\sqrt{\varepsilon D}.
\end{equation}

\section{Steady state}

\label{sec:2}

In this section we derive the asymptotics of the steady-state solution to the
Brusselator (\ref{6dec4:04}). Let
\begin{equation}
w=v+u.
\end{equation}
and rewrite (\ref{6dec4:04}) as
\begin{align}
v_{t}  &  =\delta^{2}v_{xx}+B\left(  w-v\right)  -\left(  w-v\right)  ^{2}v,\\
\frac{1}{\varepsilon}\left(  v_{t}+\tau\left(  w_{t}-v_{t}\right)  \right)
&  =Dw_{xx}-w+v+A,
\end{align}
where%
\begin{equation}
\delta^{2}=\varepsilon D.
\end{equation}
The steady-state equations then become
\begin{gather}\label{15dec2:23}
\left\{
\begin{array}
[c]{c}%
0=\delta^{2}v_{xx}+B\left(  w-v\right)  -\left(  w-v\right)  ^{2}v,\\
0=Dw_{xx}-w+v+A
\end{array}
\right.  x\in\lbrack0,L]\\
v_{x}(0)=0=v_{x}(L),\ \ \ \ \ w_{x}(0)=0=w_{x}(L).
\end{gather}
Next we expand (\ref{15dec2:23}) in $\frac{1}{D},$
\begin{align}
v  &  =v_{0}+\frac{1}{D}v_{1}+\ldots,\\
w  &  =w_{0}+\frac{1}{D}w_{1}+\ldots
\end{align}
We obtain $w_{0}^{\prime\prime}=0$ so that $w_{0}$ is a constant to be
determined. For $v_{0}$ we obtain
\begin{equation}
\delta^{2}v_{0xx}=F(v_{0},w_{0}),
\end{equation}
where
\begin{equation}
F(v,w)\equiv-B\left(  v-w\right)  +\left(  v-w\right)  ^{2}v.
\end{equation}

We seek solutions for $w_{0}$ such that $F(v_{0},w_{0})$ is a cubic in $v_{0}$
with equidistant roots. This is the so-called Maxwell line condition, and
implies that the integral of $F$ between its first two roots is the negative
of the integral between the last two. When this is the case, the solution for
$v_{0}$ will be a kink-like solution in the form of a tanh. An example of such
solution is shown on Figure \ref{fig:1}. The condition of equidistant roots is
equivalent to solving two equations
\begin{equation}
F_{vv}(v,w)=0=\ F(v,w)
\end{equation}
for unknowns $B$ and $w.$ A simple computation shows that this is equivalent
to%
\begin{equation}
B=\frac{2}{9}w_{0}^{2}.\label{Bw}%
\end{equation}
Substituting for $B$ we obtain
\begin{equation}
F(v_{0,}w_{0})=\left(  v_{0}-w_{0}\right)  ^{2}v_{0}-\frac{2}{9}w_{0}%
^{2}\left(  v_{0}-w_{0}\right)  =\left(  v_{0}-\frac{1}{3}w_{0}\right)
\left(  v_{0}-\frac{2}{3}w_{0}\right)  \left(  v_{0}-w_{0}\right)
.\label{Feq}%
\end{equation}
On the entire space, the ODE $\delta^{2}v_{0}^{\prime\prime}=F(v_{0},w_{0})$
with $F$ as in (\ref{Feq}) admits the following two solutions,
\begin{align}
v_{r} &  =\frac{2}{3}w_{0}+w_{0}\frac{1}{3}\tanh\left(  \frac{w_{0}}{3}%
\frac{x}{\sqrt{2}\delta}\right)  ,\label{vr}\\
v_{l} &  =\frac{2}{3}w_{0}-w_{0}\frac{1}{3}\tanh\left(  \frac{w_{0}}{3}%
\frac{x}{\sqrt{2}\delta}\right)  .\label{vl}%
\end{align}
We are interested in mesa-type solutions. A single, symmetric mesa-type
solution on an interval $[0,L]$ has the form,%
\begin{equation}
v_{0}\sim\left\{
\begin{array}
[c]{c}%
v_{l}\left(  x-x_{l}\right)  ,\ \ \ x<\frac{L}{2}\\
v_{r}\left(  x-x_{r}\right)  ,\ \ x>\frac{L}{2}%
\end{array}
.\right.
\end{equation}
Here, we choose%
\begin{equation}
x_{l}=\frac{L-l}{2},\ \ \ \ x_{r}=\frac{L+l}{2}.
\end{equation}
where $l$ is the width of the mesa to be determined. A K-spike symmetric
solution on the interval $\left[  0,1\right]  $ is then obtained by glueing
together $K$ solutions on the interval $L=\frac{1}{K}.$ Such a solution has
$2K$ interfaces whose locations are given by $x_{li},x_{ri}$ defined in
(\ref{14apr15:50}). To find $l$ we need the second order equations. We have,
\begin{equation}
\delta^{2}v_{1xx}=F_{v}(v_{0,}w_{0})v_{1}+F_{w}(v_{0,}w_{0})w_{1},\label{v1}%
\end{equation}%
\begin{equation}
w_{1xx}=w_{0}-v_{0}-A,\label{w1}%
\end{equation}
where
\begin{equation}
F_{v}\left(  v,w\right)  =B+\left(  v-w\right)  \left(  3v-w\right)
,\ \ \ F_{w}\left(  v,w\right)  =-B+2\left(  w-v\right)  v.
\end{equation}
Integrating (\ref{w1}) and 
using the boundary condition $w_{1}^{\prime}\left(  \pm L\right)  =0$ we
obtain
\begin{equation}
\int_{0}^{L}\left(  w_{0}-v_{0}-A\right)  =0.
\end{equation}
We evaluate
\begin{equation}
\int_{0}^{L}v_{0}\sim l\frac{w_{0}}{3}+w_{0}\left(  L-l\right)
\end{equation}
from where
\begin{equation}
l\sim\frac{3}{2}\frac{LA}{w_{0}}\sim\frac{LA}{\sqrt{2B}}.\label{lwB}%
\end{equation}
Substituting $L=\frac{1}{K}$ then yields Proposition \ref{prop:1}.

We remark that the equation (\ref{w1}) for $w_{1}$ with Neumann boundary
condition is solvable up to a constant. See Section \ref{sec:5}, Lemma
\ref{lemma:9} for the determination of this constant.

\section{Stability in the regime $DK^{2}\ll O\left(  \frac{1}{\varepsilon}%
{\ln^{2}{\varepsilon}}\right)  $}

\label{sec:3}

In this section we derive Theorem \ref{thm:largeD}, valid when $D \ll O\left(
\frac{1}{\varepsilon\ln^{2} \varepsilon}\right)  .$ Before showing this
result, we derive a more general formula for eigenvalues which is valid for
all $\tau.$ We show the following.

\begin{lemma}
\label{lemma:4} Consider a K-mesa symmetric equilibrium solution as given in
Proposition \ref{prop:1}. Moreover suppose that
\begin{equation}
1\ll DK^{2}\ll O\left(  \frac{1}{\varepsilon}{\ln^{2}{\varepsilon}}\right)  .
\end{equation}
The eigenvalues of such equilibrium state are asymptotically given implicitly
by%
\begin{equation}
\lambda\sim2\sqrt{B\frac{\varepsilon}{D}}\left(  ldK-\frac{2}{\sigma}%
\frac{\left(  \tau-1\right)  \lambda+\varepsilon}{\varepsilon}\right)
\label{12dec1:52}%
\end{equation}
where $\sigma$ may take one of the following $2K$ values%
\begin{gather}
\sigma_{j\pm}=\left(  c\pm\sqrt{a^{2}+b^{2}+2ab\cos\left(  \theta\right)
}\right)  ,\ \ \theta=\frac{\pi j}{K}\ \ \text{for}\ \ j=1\ldots K-1,\\
\sigma_{\pm}=c+a\pm b,
\end{gather}
where%
\begin{equation}
a=\frac{-\mu_{d}}{\sinh\left(  \mu_{d}d\right)  },\ \ \ \ \ \ \ b=\frac
{-\mu_{l}}{\sinh\left(  \mu_{l}l\right)  },\ \ \ \ \ \ \ c=\mu_{d}\coth\left(
\mu_{d}d\right)  +\mu_{l}\coth\left(  \mu_{l}l\right)  ,
\end{equation}%
\begin{equation}
\mu_{l}\equiv\frac{\sqrt{2\varepsilon+\lambda\left(  2\tau-1\right)  }}%
{\delta},\ \ \ \ \ \mu_{d}\equiv\frac{\sqrt{\lambda}}{\delta}.
\end{equation}

\end{lemma}

\textbf{Proof}. We start by linearizing around the equilibrium solution
$(v,w).$ We write,
\begin{equation}
v(x,t)=v\left(  x\right)  +e^{\lambda t}\phi\left(  x\right)
,\ \ \ w(x,t)=w\left(  x\right)  +e^{\lambda t}\psi\left(  x\right)
\end{equation}
where we assume that $\psi$ and $\phi$ are small. We obtain
\begin{subequations}
\begin{align}
\lambda\phi &  =\delta^{2}\phi_{xx}-F_{v}(v,w)\phi-F_{w}(v,w)\psi
,\label{eqphi}\\
\frac{1}{\varepsilon}\left(  \phi+\tau\left(  \psi-\phi\right)  \right)
\lambda &  =D\psi_{xx}-\psi+\phi, \label{eqpsi}%
\end{align}
where $\delta=\sqrt{\varepsilon D}$. Using (\ref{Bw}) we obtain
\end{subequations}
\begin{align}
F_{v}(v_{0},w_{0})  &  =3v_{0}^{2}-4w_{0}v_{0}+\left(  w_{0}^{2}+B\right)
=3v_{0}^{2}-4w_{0}v_{0}+\frac{11}{9}w_{0}^{2},\\
F_{w}(v_{0},w_{0})  &  =-2v_{0}^{2}+2w_{0}v_{0}-B=-2v_{0}^{2}+2w_{0}%
v_{0}-\frac{2}{9}w_{0}^{2}%
\end{align}
so that
\begin{align}
F_{v}(w_{0},w_{0})  &  =B,\ \ \ \ F_{w}(w_{0},w_{0})=-B,\\
F_{v}(\frac{w_{0}}{3},w_{0})  &  =B,\ \ \ \ F_{w}(\frac{w_{0}}{3},w_{0})=B.
\end{align}

Note that away from kink locations $x_{li},x_{ri},$ the diffusion term
$\varepsilon D\phi^{\prime\prime}$ may be neglected and we have $\phi
\sim-\frac{F_{w}(v_{0},w_{0})}{F_{v}(v_{0},w_{0})}\psi.$ On the other hand,
near the kink locations we locally estimate the eigenfunction by the
derivative of the profile. This suggests the following asymptotic form:%
\begin{equation}
\phi\sim\left\{
\begin{array}
[c]{l}%
c_{li}v_{li}^{\prime},\ \ \ x\sim x_{li}\\
c_{ri}v_{ri}^{\prime},\ \ \ x\sim x_{ri}\\
\psi,\ \ \ x\notin\left(  x_{li},x_{ri}\right)  ,\ \ i=1\ldots K\\
-\psi,\ \ \ x\in\left(  x_{li},x_{ri}\right)  ,\ \ i=1\ldots K
\end{array}
\right. 
\end{equation}
where the constants $c_{li}$ and $c_{ri}$ are to be found. We multiply
(\ref{eqphi}) by $v_{li}^{\prime}$ and integrate. Because of exponential
decay, we obtain that $\int v_{li}^{\prime}\phi\sim c_{li}\int v_{li}%
^{\prime2}.$ Therefore%
\begin{equation}
\lambda\int v_{li}^{\prime2}+\int v_{li}^{\prime}F_{w}(v_{li},w_{0})\psi
\sim\int\left[  \delta^{2}v_{li}^{\prime}\phi_{xx}-v_{li}^{\prime}\phi
F_{v}\left(  v_{li},w_{0}\right)  -v_{li}^{\prime}\phi\frac{1}{D}\left(
F_{vv}\left(  v_{li},w_{0}\right)  v_{1}+F_{vw}\left(  v_{li},w_{0}\right)
w_{1}\right)  \right]  .
\end{equation}
Using integration by parts we obtain,%
\begin{equation}
\int\left[  \varepsilon Dv_{li}^{\prime}\phi_{xx}-v_{li}^{\prime}\phi
F_{v}\left(  v_{li},w_{0}\right)  \right]  \sim0.
\end{equation}
Next we evaluate
\begin{equation}
I=\int v_{li}^{\prime}\phi\left(  F_{vv}\left(  v_{li},w_{0}\right)
v_{1}+F_{vw}\left(  v_{li},w_{0}\right)  w_{1}\right)  .
\end{equation}
Differentiating (\ref{v1}) we have%
\begin{equation}
\delta^{2}v_{1}^{\prime\prime}-v_{l}^{\prime}\left(  F_{vv}\left(
v_{li},w_{0}\right)  v_{1}+F_{vw}\left(  v_{li},w_{0}\right)  w_{1}\right)
-F_{v}\left(  v_{li},w_{0}\right)  v_{1}^{\prime}-F_{w}\left(  v_{li}%
,w_{0}\right)  w_{1}^{\prime}=0
\end{equation}
so that
\begin{align}
I  &  =\int\phi\left(  \delta^{2}v_{1}^{\prime\prime}-F_{v}\left(
v_{li},w_{0}\right)  v_{1}^{\prime}-F_{w}\left(  v_{li},w_{0}\right)
w_{1}^{\prime}\right) \\
&  \sim-\int c_{li}v_{l}^{\prime}F_{w}\left(  v_{li},w_{0}\right)
w_{1}^{\prime}.
\end{align}
Therefore we obtain%
\begin{equation}
c_{li}\lambda\int_{x_{li}^{-}}^{x_{li}^{+}}v_{li}^{\prime2}+\psi\left(
x_{li}\right)  \int_{x_{li}^{-}}^{x_{li}^{+}}v_{li}^{\prime}F_{w}(v_{0}%
,w_{0})\sim c_{li}w_{1}^{\prime}(x_{li})\int_{x_{li}^{-}}^{x_{li}^{+}}%
v_{li}^{\prime}F_{w}\left(  v_{0},w_{0}\right)  .
\end{equation}
Here and below, the symbol $\int_{x_{li}^{-}}^{x_{li}^{+}}$ denotes
integration over the interface located at $x_{li}$. Since $v_{0}^{\prime}$
decays exponentially outside the interface, this symbol is unambigious; that
is $\int_{x_{li}^{-}}^{x_{li}^{+}}=\int_{0}^{1}+e.s.t.$ Next we show that
\begin{equation}\label{15dec2:31}
\int_{x_{li}^{-}}^{x_{li}^{+}}v_{li}^{\prime}F_{w}\left(  v_{0},w_{0}\right)
\sim-\frac{8}{81}w_{0}^{3},\ \ \ \ \int_{x_{li}^{-}}^{x_{li}^{+}}%
v_{li}^{\prime2}\sim\frac{2\sqrt{2}w_{0}^{3}}{81\delta}.
\end{equation}
We have $\int_{x_{l}^{-}}^{x_{l}^{+}}v_{li}^{\prime}F_{w}\left(  v_{0}%
,w_{0}\right)  =G(v(x_{li}^{+}))-G(v(x_{li}^{-}))$ where $G(v)\equiv\int
F_{w}(v,w_{0})dv=-\frac{2}{3}v^{3}+w_{0}v^{2}-\frac{2}{9}w_{0}^{2}v.$ We have
$v(x_{li}^{+})=w_{0}/3,\ v(x_{li}^{-})=w_{0}$ so that $G(v(x_{li}%
^{+}))-G(v(x_{li}^{-}))=\left[  \frac{1}{81}-\frac{1}{9}\right]  w_{0}%
^{3}=-\frac{8}{81}w_{0}^{3}.$

To evaluate the second integral in (\ref{15dec2:31}), 
use the explicit formula (\ref{vl}) for
$v_{l}$ and the fact that $\int_{-\infty}^{+\infty}\operatorname{sech}%
^{4}\left(  y\right)  dy=\frac{4}{3}.$

Using (\ref{w1}) and (\ref{lwB}) yields%
\begin{equation}
w_{1}^{\prime}\left(  x_{li}\right)  \sim-\frac{w_{0}ldK}{3D}.\
\end{equation}

Therefore we obtain%
\begin{equation}
\label{23feb16:02}c_{li}\lambda\frac{\sqrt{2}}{\delta}=4\left(  c_{li}\left(
\frac{1}{3} \frac{ldK}{D}w_{0}\right)  +\psi\left(  x_{li}\right)  \right)  ,
\end{equation}
An analogous computation yields%
\begin{equation}
\label{23feb16:03}-c_{ri}\lambda\frac{\sqrt{2}}{\delta}=4\left(
-c_{ri}\left(  \frac{1}{3}\frac{ldK}{D}w_{0}\right)  +\psi\left(
x_{ri}\right)  \right)  ,
\end{equation}

It remains to compute $\ \psi\left(  x_{li}\right)  .$ Inside the intervals
$\left(  x_{li},x_{ri}\right)  $ we have $\phi\sim-\psi$ and outside those
intervals we have $\phi\sim\psi.$ In addition we assume that near kinks,
$\psi$ changes much slower than $\phi.$ In this case we may replace
\begin{equation}
v_{li}^{\prime}\sim-\frac{2}{3}w_{0}{\mbox{\boldmath$\delta$}}\left(
x-x_{li}\right)  ,\ \ v_{ri}^{\prime}\sim\frac{2}{3}w_{0}%
{\mbox{\boldmath$\delta$}}\left(  x-x_{ri}\right)
\end{equation}
inside the equation (\ref{eqpsi}). Here and below, ${\mbox{\boldmath$\delta$}}%
$ denotes the Dirac delta function. Therefore we obtain,%
\begin{equation}
\psi_{xx}-\mu^{2}\psi\sim-\alpha\sum_{i=1}^{K}c_{li}{\mbox{\boldmath$\delta$}}%
\left(  x-x_{li}\right)  -c_{ri}{\mbox{\boldmath$\delta$}}\left(
x-x_{ri}\right)
\end{equation}
where
\begin{align}\label{mul}
\mu &  \sim\mu_{l}\equiv\frac{\sqrt{2\varepsilon+\lambda\left(  2\tau
-1\right)  }}{\delta},\ \ \ x\in\left(  x_{li},x_{ri}\right)  ,\\
\label{mud}
\mu &  \sim\mu_{d}\equiv\frac{\sqrt{\lambda}}{\delta},\ \ \ x\notin\left(
x_{li},x_{ri}\right)  ,
\end{align}
and%
\begin{equation}
\alpha=\frac{2}{3}w_{0}\frac{\left(  1-\tau\right)  \lambda-\varepsilon
}{\delta^{2}}.
\end{equation}
Next we apply the following lemma.

\begin{lemma}
\label{lemma:1} Suppose that
\begin{align}
u^{\prime\prime}-\mu^{2}u  &  =-\left(  \sum b_{li}{\mbox{\boldmath$\delta$}}%
\left(  x-x_{li}\right)  +b_{ri}{\mbox{\boldmath$\delta$}}\left(
x-x_{ri}\right)  \right) \label{15dec1:49}\\
\text{with \ }u^{\prime}\left(  0\right)   &  =0=u^{\prime}\left(  1\right)  ,
\end{align}
where $x_{li},x_{ri}$ are given in (\ref{14apr15:50}) and
\begin{equation}
\mu=\left\{
\begin{array}
[c]{cc}%
\mu_{l}, & x\in\left(  x_{li},x_{ri}\right) \\
\mu_{d}, & x\notin\left(  x_{li},x_{ri}\right)
\end{array}
\right.  .
\end{equation}
Let
\begin{equation}
\vec{u}\equiv\left[
\begin{array}
[c]{c}%
u\left(  x_{l1}\right) \\
u\left(  x_{r1}\right) \\
\vdots\\
u\left(  x_{lK}\right) \\
u\left(  x_{rK}\right)
\end{array}
\right]  ,\ \ \ \ \ \ \vec{b}\equiv\left[
\begin{array}
[c]{c}%
b_{l1}\\
b_{r1}\\
\vdots\\
b_{lK}\\
b_{rK}%
\end{array}
\right]  .\ \ \ \ \ \
\end{equation}
\ \ \ \ \ \ \ \ \ \ Then
\begin{equation}
\vec{u}=\mathbf{M}\vec{b},
\end{equation}
where
\begin{equation}
M^{-1}=\left[
\begin{array}
[c]{ccccccc}%
a+c & b &  &  &  &  & \\
b & c & a &  &  &  & \\
& a & c & b &  &  & \\
&  &  & \cdots &  &  & \\
&  &  &  & a & c & b\\
&  &  &  &  & b & c+a
\end{array}
\right]
\end{equation}
with
\begin{equation}
a=\frac{-\mu_{d}}{\sinh\left(  \mu_{d}d\right)  },\ \ \ \ \ \ \ b=\frac
{-\mu_{l}}{\sinh\left(  \mu_{l}l\right)  },\ \ \ \ \ \ \ c=\mu_{d}\coth\left(
\mu_{d}d\right)  +\mu_{l}\coth\left(  \mu_{l}l\right)  .
\end{equation}
The eigenvalues $\sigma$ of $\mathbf{M}^{-1}$ are given as follows%
\begin{gather}
\sigma_{j\pm}=\left(  c\pm\sqrt{a^{2}+b^{2}+2ab\cos\left(  \theta\right)
}\right)  ,\ \ \theta=\frac{\pi j}{K}\ \ for\ \ j=1\ldots K-1\\
\sigma_{\pm}=c+a\pm b.
\end{gather}
The two eigenvalues $\sigma_{\pm}$ may be written explicitly as%
\begin{align}
\sigma_{+}  &  =\mu_{d}\tanh\left(  \mu_{d}d/2\right)  +\mu_{l}\tanh\left(
\mu_{l}l/2\right)  ,\\
\sigma_{-}  &  =\mu_{d}\tanh\left(  \mu_{d}d/2\right)  +\mu_{l}\coth\left(
\mu_{l}l/2\right)  .
\end{align}

\end{lemma}

This lemma was derived in \cite{RW}; for convenience of the reader we give its
proof in Appendix A.

Define
\begin{equation}
\vec{\psi}=\left[
\begin{array}
[c]{c}%
\psi\left(  x_{l1}\right)  \\
\psi\left(  x_{r1}\right)  \\
\vdots\\
\psi\left(  x_{lK}\right)  \\
\psi\left(  x_{rK}\right)
\end{array}
\right]  ,\ \ \ \ \ \ \ \vec{d}=\left[
\begin{array}
[c]{c}%
c_{l1}\\
-c_{r1}\\
\vdots\\
c_{lK}\\
-c_{rK}%
\end{array}
\right]  .
\end{equation}
Using Lemma \ref{lemma:1} we have,%
\begin{equation}
\vec{\psi}=\alpha\mathbf{M}\vec{d}%
\end{equation}
and we write (\ref{23feb16:02}, \ref{23feb16:03}) as%

\begin{equation}
\vec{d}\lambda\sqrt{\frac{2}{\varepsilon D}.}=4\left(  \vec{d}\left(  \frac
{1}{3}\frac{ldK}{D}w\right)  +\alpha\mathbf{M}\vec{d}\right)  .
\end{equation}
Therefore we have%
\begin{equation}
\lambda_{j}\sqrt{\frac{2}{\varepsilon D}}=4\left(  \frac{1}{3}\frac{ldK}%
{D}w+\alpha\mathbf{\sigma}^{-1}\right)  \ \ \text{for}\ \ j=1\ldots2K
\end{equation}
where $\sigma$ are $2K$ eigenvalues of $\mathbf{M}^{-1}\mathbf{.}$

This completes the proof of Lemma \ref{lemma:4}. $\blacksquare$

We now come back to the proof of Theorem \ref{thm:largeD}. Assuming
$O(\tau-1)\gg0,$ the dimensional analysis shows that two scalings are
possible, either $\lambda=O(\varepsilon)$ or $\lambda=O(D\varepsilon).$

We first consider the case
\begin{equation}
\lambda=\varepsilon\lambda_{0}.
\end{equation}
Using the notation of Lemma \ref{lemma:4} we then obtain from (\ref{mul}) and
(\ref{mud})
\begin{equation}
\mu_{d}^{2}=\frac{\lambda_{0}}{D}\ll1,\ \ \ \ \mu_{l}^{2}=\frac{2+\lambda
_{0}\left(  2\tau-1\right)  }{D}\ll1.
\end{equation}
and
\begin{equation}
a\sim-\frac{1}{d},\ \ b\sim-\frac{1}{l},\ \ \ \ c\sim\frac{1}{d}+\frac{1}%
{l}=\frac{1}{Kdl},
\end{equation}%
\begin{align}
\sigma_{j\pm}  &  =c\pm\sqrt{\left(  a+b\right)  ^{2}-2abt}%
,\ \ \ \ \ \ \ \ t=1-\cos\left(  \frac{\pi j}{K}\right)  \in\left(  0,2\right)
\\
&  =\frac{1}{Kdl}\pm\sqrt{\left(  \frac{1}{Kdl}\right)  ^{2}-2\frac{1}{dl}t}\\
&  =\frac{1}{Kdl}\left(  1\pm\sqrt{1-2K^{2}dlt}\right)  .
\end{align}
Therefore we obtain%
\begin{equation}
\lambda_{0}\sim2ldK\sqrt{\frac{B}{\varepsilon D}}\left(  1-2\frac{\left(
\tau-1\right)  \lambda_{0}+1}{1\pm\sqrt{1-2K^{2}dlt}}\right)  .
\end{equation}
When $\lambda_{0}=O\left(  1\right)  ,$ the right hand side dominates and we
therefore obtain%
\begin{equation}
\lambda_{0}=\frac{-1\pm\sqrt{1-2K^{2}dlt}}{2\left(  \tau-1\right)  }.
\label{8mar18:01}%
\end{equation}
Moreover, since $d=\frac{1}{K}-l,\ \ l\in\left(  0,\frac{1}{K}\right)  ,$ and
since $t\in(0,2)$ we see that $2K^{2}dlt\leq1.$ This shows that in the case
$\tau>1,$ the eigenvalues corresponding to the nodes $\sigma_{j\pm}$ are all
real negative. The node $\sigma_{+}$ is,%

\begin{align}
\sigma_{+}  &  =\mu_{d}\tanh\left(  \mu_{d}\frac{d}{2}\right)  +\mu_{l}%
\tanh\left(  \mu_{l}\frac{l}{2}\right) \\
&  \sim\mu_{d}^{2}\frac{d}{2}+\mu_{l}^{2}\frac{l}{2}\\
&  \sim\frac{1}{2D}\left(  \lambda_{0}d+\left(  2+\lambda_{0}\left(
2\tau-1\right)  \right)  l\right)  ,\\
\lambda_{0}  &  \sim2\sqrt{\frac{B}{\varepsilon D}}\left(  ldK-\frac{4D\left[
\left(  \tau-1\right)  \lambda_{0}+1\right]  }{\left(  \lambda_{0}d+\left(
2+\lambda_{0}\left(  2\tau-1\right)  \right)  l\right)  }\right)  .
\end{align}
Since the third term is asymptotically bigger, the assumption $\lambda
_{0}=O(1)$ leads to%
\begin{equation}
\lambda_{0}=\frac{-1}{\tau-1}.
\end{equation}
Note that this is a special case of the formula (\ref{8mar18:01})\ with
$\pm=-$ and $t=0.$ For the mode $\sigma_{-}$ we have%
\begin{equation}
\sigma_{-}=\mu_{d}\tanh\left(  \mu_{d}\frac{d}{2}\right)  +\mu_{l}\coth\left(
\mu_{l}\frac{l}{2}\right)  \sim\frac{2}{l};
\end{equation}%
\begin{align}
\lambda_{0}  &  \sim2l\sqrt{B\frac{1}{\varepsilon D}}\left(  dK-\left[
\left(  \tau-1\right)  \lambda_{0}+1\right]  \right)  ,\\
\lambda_{0}  &  \sim\frac{-1+Kd}{\tau-1}\sim\frac{-Kl}{\tau-1}.
\end{align}
Note that this is a special case of the formula (\ref{8mar18:01})\ with
$\pm=-$ and $t=2.$

Next we consider large eigenvalues, $\lambda\gg O(\varepsilon).$ Then we have
\begin{equation}
\mu_{l}\sim\frac{\sqrt{\left(  2\tau-1\right)  \lambda}}{\sqrt{D\varepsilon}%
},\ \ \ \mu_{d}\sim\frac{\sqrt{\lambda}}{\sqrt{D\varepsilon}},
\end{equation}
and we write,
\begin{equation}
\lambda\sim2\sqrt{B}\left(  \frac{1}{D}\sqrt{D\varepsilon}ldK-\frac{2}{\left(
\sigma/\frac{\sqrt{\lambda}}{\sqrt{D\varepsilon}}\right)  }\left(
\tau-1\right)  \sqrt{\lambda}\right)  .
\end{equation}
Dimensional analysis shows that the only way to achieve this balance is when
$\sigma$ is large. But this is only possible when%
\begin{equation}
\sinh\left(  \mu_{l}l\right)  =0\text{ \ \ or \ \ \ }\sinh\left(  \mu
_{d}d\right)  =0.
\end{equation}
Thus either $\mu_{l}l=i\pi m$ or $\mu_{d}d=i\pi m$ where $m$ is some integer.
This yields the following eigenvalues,%
\begin{equation}
\lambda\sim-D\varepsilon\frac{m^{2}\pi^{2}}{l^{2}\left(  2\tau-1\right)
}\text{ \ \ \ \ or \ \ }\lambda\sim-D\varepsilon\frac{m^{2}\pi^{2}}{d^{2}}.
\end{equation}

Finally, we show that a Hopf bifurcation occurs in the regime $O\left(
\tau-1\right)  \ll1.$ Since the small eigenvalues are negative when $\tau
-1\gg0$ and positive when $\tau-1\ll0$, the real part changes sign precisely
when $O\left(  \tau-1\right)  \ll1.$ \ To show that this occurs via a\ Hopf
bifurcation, it suffices to show that $\lambda$ can never be zero. Suppose
not. Then using some algebra we arrive at the following for the modes
$\sigma_{j\pm}$%
\begin{equation}
-1\pm\sqrt{1-2K^{2}dlt}=0,\text{ \ }\ t=1-\cos\left(  \frac{\pi j}{K}\right)
\in\left(  0,2\right)  .
\end{equation}
But this is clearly impossible since $K^{2}dlt<\frac{1}{4}$ as mentioned
above. The modes $\sigma_{\pm}$ are handled similarly.

This completes the proof of Theorem \ref{thm:largeD}. $\blacksquare$

\section{\bigskip Hopf bifurcation, $DK^{2}\ll O\left(  \frac{1}{\varepsilon
}{\ln^{2}{\varepsilon}}\right)  $}

\label{sec:4}

As Theorem \ref{thm:largeD} shows, a Hopf bifurcation occurs when $\tau$ is
near 1. In this section we study this regime in more detail, culminating in
the proof of Theorem \ref{thm:hopf}.

We start by analysing the $\sigma_{+}$ node. We make the assumption that%
\begin{equation}
\lambda\ll O\left(  \varepsilon D\right)  .
\end{equation}
This assumption will be verified later on to be consistent with the condition
(\ref{9mar14:03}), $D\gg\sqrt{\frac{B}{\varepsilon D}}$. We have,
\begin{equation}
\mu_{l}^{2}\sim\frac{\lambda+2\varepsilon}{\varepsilon D}\ll1,\ \ \ \mu
_{d}^{2}\sim\frac{\lambda}{\varepsilon D}\ll1,
\end{equation}
and we expand $\sigma_{+}$ up to second order,%
\begin{align}
\sigma_{+}  &  =\mu_{d}\tanh\left(  \mu_{d}\frac{d}{2}\right)  +\mu_{l}%
\tanh\left(  \mu_{l}\frac{l}{2}\right) \\
&  \sim\frac{\lambda}{\varepsilon D}\frac{d}{2}-\frac{1}{24}\left(
\frac{\lambda}{\varepsilon D}\right)  ^{2}d^{3}+\frac{\lambda+2\varepsilon
}{\varepsilon D}\frac{l}{2}-\frac{1}{24}\left(  \frac{\lambda+2\varepsilon
}{\varepsilon D}\right)  ^{2}l^{3}\\
&  \sim-\left(  \frac{\lambda}{\varepsilon D}\right)  ^{2}\left(  \frac
{d^{3}+l^{3}}{24}\right)  +\frac{\lambda}{\varepsilon D}\left(  \frac{1}%
{2K}\right)  +\frac{l}{D}.
\end{align}
We write the equation for $\lambda$ as%
\begin{equation}
\lambda\sigma\sim2\sqrt{B\frac{\varepsilon}{D}}\left(  ldK\sigma
-2\frac{\left(  \tau-1\right)  \lambda+\varepsilon}{\varepsilon}\right)
\end{equation}
or%
\begin{equation}
a\lambda^{3}+b\lambda^{2}+c\lambda+e=0,
\end{equation}
where%
\begin{align}
a  &  =\left(  \frac{1}{\varepsilon D}\right)  ^{2}\left(  \frac{d^{3}+l^{3}%
}{24}\right)  ,\\
b  &  =-\frac{1}{\varepsilon D}\left(  \frac{1}{2K}\right)  -2\sqrt
{B\frac{\varepsilon}{D}}ldK\left(  \frac{1}{\varepsilon D}\right)  ^{2}\left(
\frac{d^{3}+l^{3}}{24}\right) \\
&  \sim-\frac{1}{\varepsilon D}\left(  \frac{1}{2K}\right)  \ \ \text{using
(\ref{9mar14:03});}\\
c  &  =-\frac{l}{D}+2\sqrt{B\frac{\varepsilon}{D}}\left(  \frac{ld}%
{2\varepsilon D}-\frac{2}{\varepsilon}\left(  \tau-1\right)  \right) \\
&  \sim2\sqrt{B\frac{\varepsilon}{D}}\left(  \frac{ld}{2\varepsilon D}%
-\frac{2}{\varepsilon}\left(  \tau-1\right)  \right)  ;\\
e  &  =2\sqrt{B\frac{\varepsilon}{D}}\left(  ldK\left\{  \frac{l}{D}\right\}
-2\right) \\
&  \sim-4\sqrt{B\frac{\varepsilon}{D}}.
\end{align}
Substituting $\lambda=i\lambda_{i}$ and separating the real and imaginary
part, we find that%
\begin{equation}
\lambda_{i}=\sqrt{\frac{e}{b}},\ \ \ c=\frac{ae}{b}.
\end{equation}
This yields,%
\begin{equation}
\lambda_{i}=\sqrt{8K}\left(  \varepsilon^{3}DB\right)  ^{1/4}%
\end{equation}
and, keeping also all lower order terms for reference,%
\begin{equation}
\tau-1\sim\frac{ld}{4D}-\frac{\left(  \frac{d^{3}+l^{3}}{12}\right)
K}{D+\sqrt{B\frac{1}{\varepsilon D}}ldK^{2}\left(  \frac{d^{3}+l^{3}}%
{6}\right)  }-\frac{l\varepsilon}{4\sqrt{B\varepsilon D}}. \label{9mar14:15}%
\end{equation}
The formula (\ref{9mar14:14}) is obtained by dropping the last term of the
right hand side of (\ref{9mar14:15}) (which is of smaller order than the first
term on the right hand side), as well as dropping the second term in the
denominator of the second term.

Finally we must also show that the $\sigma_{+}$ mode undergoes a Hopf
bifurcation before all other modes.\ Let us now consider the $\sigma_{-}$
mode. \ We assume here that $\mu_{l}\sim\frac{\sqrt{\lambda}}{\sqrt
{\varepsilon D}}\sim\mu_{d}.$ and we rewrite (\ref{12dec1:52}) as
\begin{align}\label{15dec2:43}
-4\frac{1}{D}\sqrt{\frac{B}{\varepsilon D}}\left(  \tau_{0}\lambda
_{0}D+1\right)   &  =F(\lambda_{0})\equiv\lambda_{0}\sqrt{\lambda_{0}}\left(
\tanh\sqrt{\lambda_{0}}\frac{d}{2}+\coth\sqrt{\lambda_{0}}\frac{l}{2}\right)
\\
\text{where }\tau &  =1+\tau_{0},\text{ \ }\lambda=\varepsilon D\lambda_{0}.
\end{align}
Near the origin and for real $\lambda_{0}$, the curve $F\left(  \lambda
_{0}\right)  \sim\frac{2}{l}\lambda_{0}+\left(  \frac{d}{2}+\frac{l}%
{6}\right)  \lambda_{0}^{2}$ is convex and increasing. The left hand side 
of (\ref{15dec2:43}) is a
line in $\lambda_{0}$ and it intersects the y axis at $-\frac{4}{D}\sqrt
{\frac{B}{\varepsilon D}}$ which is a very small value by assumption
(\ref{9mar14:03}). Therefore this line will intersect the curve $F\left(
\lambda_{0}\right)  $ for some small value of $\lambda_{0}$ unless its slope
is precisely the slope of $F(\lambda_{0})$ at the origin, i.e. $-4\sqrt
{\frac{B}{\varepsilon D}}\tau_{0}\sim\frac{2}{l}.$ This is precisely the
scaling on which the Hopf bifurcation occurs. Subsituting $l=\frac{A}%
{\sqrt{2B}}$ we obtain,%
\begin{equation}
\tau_{h_{-}}\sim1-\frac{\sqrt{\varepsilon D}}{\sqrt{2}A}. \label{10mar18:11}%
\end{equation}
Performing a similar study for the modes $\sigma_{j\pm}$ we obtain
\begin{equation}
\tau_{h_{j\pm}}\sim1-\frac{\sqrt{\varepsilon D}}{\sqrt{B}4Kdl}\left(
1\pm\sqrt{1-2K^{2}dlt}\right)  ,\ \ \ \ t=\ 1-\cos\left(  \frac{\pi j}%
{K}\right)  \in\left(  0,2\right)  . \label{10mar18:12}%
\end{equation}
But clearly, $\tau_{h_{j\pm}},\tau_{h_{-}}<\tau_{h_{+}}$ since $\frac
{\sqrt{\varepsilon D}}{A}\gg O\left(  \frac{1}{D}\right)  $ by the assumption
(\ref{9mar14:03})$.$ This shows that the eigenvalue $\lambda_{+}$
corresponding to $\sigma_{+}$ undergoes the Hopf bifurcation before any of the
other eigenvalues, as $\tau$ decreases past $\tau_{h_{+}}$. $\blacksquare$

\begin{figure}[ptb]
\begin{center}
\setlength{\unitlength}{1.75cm} \begin{picture}(5.0,3.1)(0,0)
\put(2.5, 0){$\log_{10}(D \delta)$}
\put(-0.05, 1.2){\rotatebox{90}{$\log_{10}(1-\tau)$}}
\put(0.2,0.2){\includegraphics{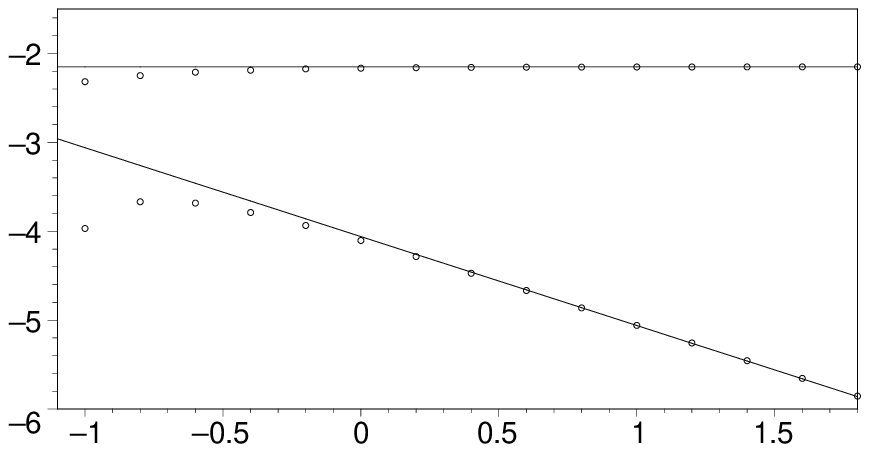}}
\end{picture}
\end{center}
\caption{The value of $\tau_{h_{+}}$ and $\tau_{h_{-}}$ as a function of $D$,
while $\delta=\sqrt{\varepsilon D}=0.01$ is held fixed. The dots represent the
numerical solution obtained by substituting $\lambda=i \lambda_{i}$ into
(\ref{12dec1:52}) and solving for $\tau$ and $\lambda_{i}$ using Newton's
method. The solid lines are represent formulas (\ref{9mar14:14}) and
(\ref{10mar18:11}). Here, $A=1$ and $B=15.$ }%
\label{fig:tauhopf}%
\end{figure}

In figure \ref{fig:tauhopf} we show the Hopf bifurcation values $\tau_{h_{+}}$
and $\tau_{h_{-}}$ computed numerically as well as the asymptotic results
(\ref{9mar14:14}), (\ref{10mar18:11}), for various values of $D$ while fixing
$\delta=\sqrt{D\varepsilon}=0.01.$ This figure shows a very good agreement
when $D\gg\delta.$

\section{\bigskip Asymmetric K-mesa solutions and instability with $DK^{2}\sim
O\left(  \frac{1}{\varepsilon}{\ln^{2}{\varepsilon}}\right)  $}

\label{sec:5}

In Section \ref{sec:3} we have shown that $K$ mesas are always stable provided
that $\tau> 1$. In our analysis there, we have ignored the effect of the
exponentially decaying tail of $v$. However as $D K^{2}$ increases, this
effect eventually must be taken into account. As we will see in this section,
this occurs when $D\geq O\left(  \frac{1}{\varepsilon\ln^{2}\varepsilon
}\right)  .$ The main result of this section is the following.

\begin{proposition}
\label{prop:expstab} Let
\begin{equation}
f(D)=3\sqrt{B/2}+\frac{A}{16BD}\left(  \sqrt{2B}-A\right)  ^{2}+3\sqrt
{2B}\left(  \exp\left\{  -\frac{A}{\sqrt{2\varepsilon D}}\right\}
+\exp\left\{  -\frac{1}{\sqrt{2\varepsilon D}}\left(  \sqrt{2B}-A\right)
\right\}  \right)  ,
\end{equation}
and let $D_{1}$ be the minimum of $f(D)$. Let%
\begin{equation}
D_{K}=\frac{1}{K^{2}}D_{1}.
\end{equation}
Suppose that $\tau>1$ and $K\geq2.$ Then $K-mesa$ solution of Proposition
\ref{prop:1} is stable when $D<D_{K}$ and unstable when $D>D_{K}.$ The minimum
$D_{1}$ satisfies the following transcendental equation,%
\begin{equation}
\ \ \ \ \ \frac{A}{\sqrt{D_{1}}8B}\left(  \sqrt{2B}-A\right)  ^{2}%
=\frac{3\sqrt{2B}}{\sqrt{2\varepsilon}}\left(  A\exp\left\{  -\frac{A}%
{\sqrt{D_{1}2\varepsilon}}\right\}  +\left(  \sqrt{2B}-A\right)  \exp\left\{
-\frac{1}{\sqrt{D_{1}2\varepsilon}}\left(  \sqrt{2B}-A\right)  \right\}
\right)  . \label{D1}%
\end{equation}

Suppose that $2A^{2}<B.$ Then%
\begin{equation}
D_{1}\sim\frac{A^{2}}{2\varepsilon\ln^{2}\left(  \frac{12\sqrt{2}AB^{3/2}%
}{\varepsilon\left(  \sqrt{2B}-A\right)  ^{2}}\right)  }.
\end{equation}
Suppose that $2A^{2}>B.$ Then%
\begin{equation}
D_{1}\sim\frac{\left(  \sqrt{2B}-A\right)  ^{2}}{2\varepsilon\ln^{2}\left(
\frac{12\sqrt{2}}{\varepsilon A}B^{3/2}\right)  }.
\end{equation}

\end{proposition}

\begin{figure}[ptbh]
\begin{center}
\setlength{\unitlength}{1in} \begin{picture}(6.5,2.4)(0.4,0)
\put(0.16,0.25){\includegraphics{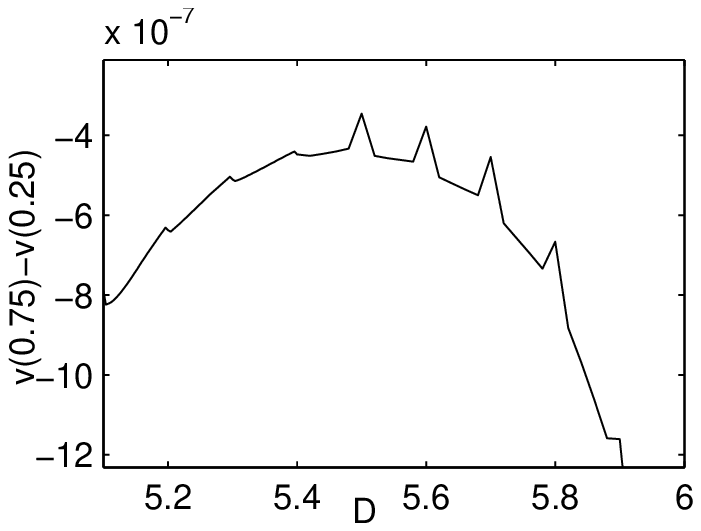}}
\put(3.333,0.25){\includegraphics{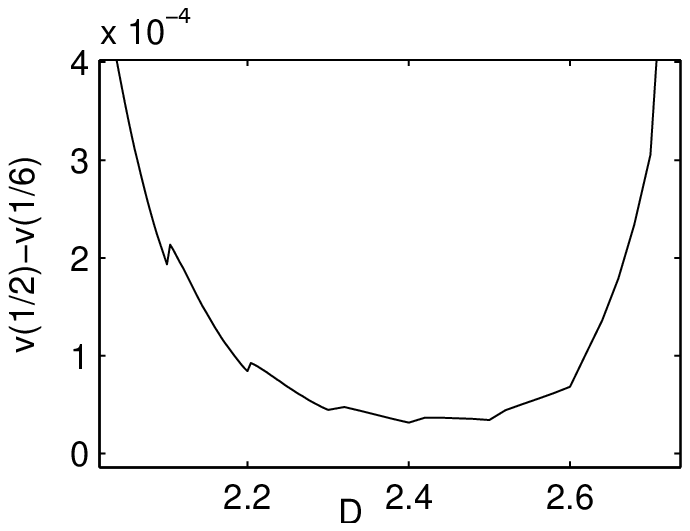}}
\put(1.5,0.1){(a)}
\put(4.8,0.1){(b)}
\end{picture}
\end{center}
\caption{ (a) The difference in height of a two-mesa solution for various
values of $D$. Here, $D=5+0.1\text{floor}(t/2500)$. Note the change of
stability when $D \sim5.5$. (b) The difference in height of the first two
mesas of a three-mesa solution. Here, $D=1.9+0.1\text{floor}(t/2500)$. Note
the change of stability when $D \sim2.45$. In both figures, $\varepsilon
=0.001,\ A=2,\ B=18, \tau=3.$ }%
\label{fig:K2diff}%
\end{figure}

Before providing a proof, consider a numerical example. Take
\begin{equation}
\varepsilon=0.001,\ A=2,\ B=18.
\end{equation}
Then solving (\ref{D1}) we obtain $D_{1}=21.16$ so that
\begin{equation}
D_{2}=5.3,\ \ \ D_{3}=2.35.
\end{equation}
To verify Proposition \ref{prop:expstab}, we ran the full numerical simulation
of (\ref{6dec4:04}) for various values of $D$. We took $\tau=3,$ the initial
condition to be as given in Proposition \ref{prop:1} with $K=2,$ and we took
$D$ from 5 to 6 with 0.1 increments every 2500 time units. For each value of
$D$, we then plotted the difference of the height of the two mesas versus
time. See Figure \ref{fig:K2diff}.a. From this computation, we see that a
change of stability occurs when $D\approx5.5$: if $D<5.5$ then the difference
in height is decreasing but is increasing if $D>5.5$. This agrees well with
the theoretical prediction $D=5.3.$ We then took $K=3$, and $D$ from 1.9 to 3
with 0.1 increments every 2500 time units. Figure \ref{fig:K2diff}.b shows the
the difference in heights of the first and second mesa. From this figure we
conclude that the change in stability occurs when $D\sim2.45.$ Again, this
agrees well with the theoretical prediction of $D_{3}=2.35$.

Proposition \ref{prop:expstab} follows from the existence of asymmetric
patterns. Indeed a similar phenomena was studied for the spike solutions of
the Gierer-Meinhardt model \cite{IWW} and in the Gray-Scott model \cite{KWW}.
In both of these models, an asymmetric spike pattern was found to bifurcate
from a symmetric $K$ spike solution when $K>1$. Moreover a change of stability
of a $K$-spike pattern occured precisely at that point.

To show existence of asymmetric patterns, it suffices to compute $w(L)$ as a
function of the domain length $L$, and to show an existence of a minimum of
this curve. Below we show that such minimum occurs precisely when the the
interaction in the $u$ component is balanced by the interaction of $v$ in the
exponential tail. The main result is the following.

\begin{lemma}
\label{lemma:9} Consider a symmetric mesa-type solution on domain $[0,L].$
Then we have,%
\begin{equation}
w\left(  L\right)  \sim3\sqrt{B/2}+\frac{1}{D}\frac{A}{16B}L^{2}\left(
\sqrt{2B}-A\right)  ^{2}+3\sqrt{2B}\left(  \exp\left\{  -\frac{LA}%
{\sqrt{2\varepsilon D}}\right\}  +\exp\left\{  -\frac{L}{\sqrt{2\varepsilon
D}}\left(  \sqrt{2B}-A\right)  \right\}  \right)  .\label{w(L)}%
\end{equation}

\end{lemma}

\begin{figure}[ptbh]
\begin{center}
\setlength{\unitlength}{1in} \begin{picture}(5.2,3.1)(0,0)
\put(0.1,0.0){\includegraphics[width=5in, height=3in]{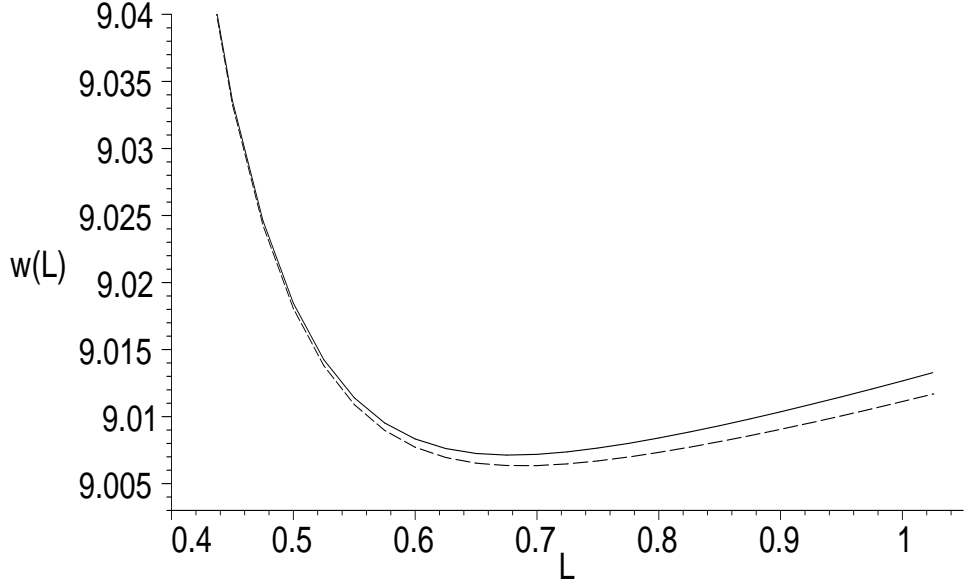}}
\end{picture}
\end{center}
\caption{The value of $w(L)$ as a function of $L$. Here,
$A=2,\ B=18,\ \varepsilon=0.001$ and $D=10.$ The solid curve represents an
exact numerical value computed using the boundary value problem solver; the
dashed curve represents the asymptotic formula (\ref{w(L)}). Note that both
curves give almost the same minimum value of $L \approx0.68$}%
\label{fig:5}%
\end{figure}

\textbf{Proof}. We recall that upon expanding the solution in $\frac{1}{D}$ as
$w=w_{0}+\frac{1}{D}w_{1}+\ldots$, $v=v_{0}+\frac{1}{D}v_{1}+\ldots$,
the\ equation for $w_{1}$ is%
\begin{equation}
\delta^{2}v_{1xx}=F_{v}(v_{0,}w_{0})v_{1}+F_{w}(v_{0,}w_{0})w_{1}%
,\label{14dec8:32}%
\end{equation}%
\begin{equation}
w_{1xx}=w_{0}-v_{0}-A.
\end{equation}

\bigskip

\bigskip Note that we have%
\begin{equation}
\delta^{2}v_{0xx}^{\prime}=F_{v}(v_{0},w_{0})v_{0}^{\prime}.
\end{equation}
Therefore, upon multiplying (\ref{14dec8:32}) by $v_{0}^{\prime}$ and
integrating we obtain,%
\begin{equation}
\int_{0}^{L/2}v_{0}^{\prime}F_{w}(v_{0}w_{0})w_{1}=\delta^{2}\left(
v_{1x}v_{0x}-v_{1}v_{0xx}\right)  _{x=0}^{x=L/2}. \label{21feb15:44}%
\end{equation}
To evaluate the right hand side, we write
\begin{equation}
\int_{0}^{L/2}v_{0}^{\prime}F_{w}(v_{0}w_{0})w_{1}\sim w_{1}\left(
x_{l}\right)  \int_{x_{l}^{-}}^{x_{l}^{+}}\frac{d}{dx}G\left(  v_{0}%
,w_{0}\right)  ,
\end{equation}
where
\begin{gather}
G(v_{0})=\int F_{w}dv_{0}=-Bv_{0}+v_{0}^{2}w_{0}-\frac{2}{3}v_{0}^{3},\\
G(w_{0})=\frac{1}{9}w_{0}^{3},\ \ \ \ G(\frac{w_{0}}{3})=\frac{1}{81}w_{0}%
^{3}.
\end{gather}
It follows that
\begin{equation}
\int_{0}^{L/2}v_{0}^{\prime}F_{w}(v_{0}w_{0})w_{1}\sim-w_{1}\left(
x_{l}\right)  \frac{8}{81}w_{0}^{3}.
\end{equation}
To evaluate the right hand side of (\ref{21feb15:44}) we expand the solution
near the boundary.

At $x=0,$ we assumed that $v\sim w_{0};$ writing
\begin{equation}
v=w_{0}+V(x)
\end{equation}
we then obtain to leading order,%
\begin{align}
\delta^{2}V^{\prime\prime}  &  =F_{v}\left(  w_{0},w_{0}\right)  V+O\left(
\frac{1}{D}\right) \\
&  \sim BV.
\end{align}
Imposing the boundary condition $V^{\prime}\left(  0\right)  =0$ we then
obtain%
\begin{equation}
V\sim K\left(  \exp\left\{  -\frac{\sqrt{B}}{\delta}x\right\}  +\exp\left\{
+\frac{\sqrt{B}}{\delta}x\right\}  \right)  +O\left(  \frac{1}{D}\right)
\end{equation}
for some constant $K.$ To determine $K$, we impose the matching condition
$w_{0}+V\sim v_{0}+\frac{1}{D}v_{1}$ in the region $\frac{\delta}{\sqrt{B}}\ll
x\ll x_{l}.$ In this region we obtain%
\begin{align}
v_{0}  &  =\frac{2}{3}w_{0}-\frac{w_{0}}{3}\tanh\frac{x-x_{l}}{2}\frac
{\sqrt{B}}{\delta}\\
&  \sim w_{0}+\frac{2}{3}w_{0}\exp\left\{  -\frac{\sqrt{B}}{\delta}%
x_{l}\right\}  \exp\left\{  \frac{\sqrt{B}}{\delta}x\right\}  .
\end{align}
Thus we obtain%
\begin{equation}
K=\frac{2}{3}w_{0}\exp\left\{  -\frac{\sqrt{B}}{\delta}x_{l}\right\}  .
\end{equation}
Moreover, for $x\ll x_{l}$ we have%
\begin{align}
\delta^{2}v_{1xx}  &  \sim Bv_{1}-Bw_{1}\\
&  \sim Bv_{1}-Bw_{1}(0)
\end{align}
so that
\begin{equation}
v_{1}\sim w_{1}\left(  0\right)  +DK\exp\left\{  -\frac{\sqrt{B}}{\delta
}x\right\}  .
\end{equation}
It follows that
\begin{align}
v_{1x}\left(  0\right)  v_{0x}\left(  0\right)   &  \sim\frac{-B}{\delta^{2}%
}K^{2}D,\ \ \ \ \ v_{1x}\left(  0\right)  v_{0xx}\left(  0\right)  \ \sim
\frac{B}{\delta^{2}}\frac{K^{2}}{D}\ ,\\
\delta^{2}\left(  v_{1x}v_{0x}-v_{1}v_{0xx}\right)  _{x=0}  &  =-4B^{2}%
D\exp\left\{  -\frac{\sqrt{B}}{\delta}\left(  L-l\right)  \right\}  .
\end{align}
Similarly, near $x=\frac{L}{2}$ we have%
\begin{equation}
v_{0}\sim\frac{w_{0}}{3}+\frac{2}{3}w_{0}\exp\left\{  \frac{-\sqrt{B}}{\delta
}\frac{l}{2}\right\}  \exp\left\{  \frac{-\sqrt{B}}{\delta}\left(  x-\frac
{L}{2}\right)  \right\}
\end{equation}
from where we deduce
\begin{equation}
v_{1}\sim-w_{1}(0)+D\frac{2}{3}w_{0}\exp\left\{  \frac{-\sqrt{B}}{\delta}%
\frac{l}{2}\right\}  \exp\left(  \frac{\sqrt{B}}{\delta}\left(  x-\frac{L}%
{2}\right)  \right)  ,
\end{equation}

\begin{equation}
\delta^{2}\left(  v_{1x}v_{0x}-v_{1}v_{0xx}\right)  _{x=L/2}=-\frac{4B^{2}}%
{D}\exp\left\{  -\frac{\sqrt{B}}{\delta}l\right\}  .
\end{equation}
Therefore we obtain%
\begin{align}
w_{1}\left(  x_{l}\right)  \frac{8}{81}w_{0}^{3}  &  \sim4B^{2}D\left(
\exp\left\{  -\frac{\sqrt{B}}{\delta}l\right\}  +\exp\left\{  -\frac{\sqrt{B}%
}{\delta}\left(  L-l\right)  \right\}  \right) \\
w_{1}\left(  x_{l}\right)   &  \sim3\sqrt{2B}D\left(  \exp\left\{  -\frac
{LA}{\sqrt{2}\delta}\right\}  +\exp\left\{  -\frac{L}{\sqrt{2}\delta}\left(
\sqrt{2B}-A\right)  \right\}  \right)  .
\end{align}
In the region $0<x<x_{l}$, we have $v_{0}\sim w_{0}$ so from (\ref{w1}), \ we
obtain
\begin{equation}
w_{1}^{\prime\prime}\sim-A,\text{ \ \ \ \ \ }0<x<x_{l}.
\end{equation}
It follows that%
\begin{equation}
w_{1}\left(  x_{l}\right)  -w_{1}\left(  0\right)  \sim-\frac{A}{2}x_{l}%
^{2}=-\frac{A}{2}\left(  \frac{L-l}{2}\right)  ^{2}.
\end{equation}
Using $w_{0}=3\sqrt{B/2}$ and\ $l=LA/\sqrt{2B},$ we then obtain (\ref{w(L)}).
$\blacksquare$

In Figure \ref{fig:5} we compare the asymptotic formula (\ref{w(L)}) with the
numerically computed value for $A=2,\ B=18,\ \varepsilon=0.001$ and $D=10.$
Note that the function $L\rightarrow w(L)$ has a minimum at $L\approx0.7.$
This shows the existence of a fold point. Now suppose that $D$ is chosen such
that this minimum occurs precisely at $L=\frac{1}{K},$ with $K>1$ an integer.
Then the corresponding $K$-mesa equilibrium solution will have a zero
eigenvalue. Since the exponential terms in (\ref{w(L)}) quickly die out as $L$
increases, the solution becomes stable to the right of $L=\frac{1}{K},$ and
therefore unstable to the left of it. Proposition \ref{prop:expstab} is
precisely this statement; it is obtained simply by scaling the $L$ out and
stating the existence of the fold point in terms of $D$ instead.

\section{Turing analysis}

\label{sec:6}

In this section we perform a Turing analysis of the homogenous steady state
$u=A,\ v=\frac{B}{A}.$ In particular, we are interested in examining any
possible connections between the Turing instability regime (which leads to
cosinusoidal-like patterns $\cos\left(  2k\pi x\right)  $ of mode $k$) and the
localized mesa-like structures. We start by linearizing (\ref{6dec4:04}%
) around the steady state as follows,%
\begin{equation}
u=A+\xi e^{\lambda t}\cos\left(  2k\pi x\right)  ,\ \ v=\frac{B}{A}+\eta
e^{\lambda t}\cos\left(  2k\pi x\right)  ,\ \ \ \ \ \xi,\eta\ll1;\ \ \ 2k\in%
\mathbb{N}
.
\end{equation}
This yields a 2x2 eigenvalue problem for $\lambda.$ Its solution is given by%
\begin{equation}
\lambda^{2}-T\lambda+\Delta=0,
\end{equation}
where%
\begin{equation}
T=B-\tau A^{2}-n\left(  1+\tau\right)  -\varepsilon,\ \ \ \ \Delta=\tau\left[
n\left(  n-B+A^{2}\right)  +\varepsilon\left(  A^{2}+n\right)  \right]
,\ \ \ \ \ n=4k^{2}\pi^{2}\varepsilon D.
\end{equation}

Note that $\Delta_{n=0}>0$ so that the zero mode is unstable if $T_{n=0}>0$ or
$B-\tau A^{2}>0.$ Numerically, we observe that when the zero mode is unstable,
it dominates and the system moves away from the equilibrium and quickly
approaches a very long relaxation cycle, before any of the non-zero modes are
activated. Therefore no spatial instability is observed. This leads to the
following \emph{necessary condition }for the Turing instability to appear:%
\begin{equation}
B-\tau A^{2}<0. \label{11mar11:03}%
\end{equation}
Provided this condition is satisfied, we have $T<0$ for all $n.$ Therefore
Turing instability will occur iff $\Delta<0.$ In particular the second
necessary condition is that
\begin{equation}
B-A^{2}>0. \label{11mar11:04}%
\end{equation}
In this case, the most unstable mode is of the same order as the minimum of
$\Delta$,
\begin{align}
\label{15apr17:15}k_{\ast}^{2}\sim\frac{B-A^{2}-\varepsilon}{\pi
^{2}2\varepsilon D}.
\end{align}
Shortly after the Turing instability is triggered, localized mesa-type
structures appear due to the presence of steep gradients. In this regime,
Turing instability cannot predict the final number $K$ of mesas. Indeed, we
have $K\leq K_{\ast}$ where%
\begin{equation}
K_{\ast}=\max\left(  1,\sqrt{\frac{D_{1}}{D}}\right)
\end{equation}
with $D_{1}=O\left(  \frac{1}{\varepsilon\ln^{2}\varepsilon}\right)  $ as
obtained in Theorem \ref{thm:hugeD}. It follows that as long as $B-A^{2}\gg0$
and $\tau-1\gg0$, we have
\begin{equation}
K_{\ast} = O\left(  \frac{1}{\delta\ln\frac{1}{\varepsilon}}\right)  ,~~
k_{\ast} = O\left(  \frac{1}{\delta} \right)  , ~~~\delta= \sqrt{\varepsilon
D}
\end{equation}
so that $K_{\ast} \ll k_{\ast}.$ Therefore we expect that shortly after the
patterns appear, a coarsening process takes place whereby some of the
resulting mesas dissapear until there are at most $K_{\ast}$ of them left.

Consider an example shown on Figure \ref{fig:turing}.a. Take $A=1,\ B=8,$
$\varepsilon=10^{-4},\ D=10.$ We take $\tau=10$ to satisfy (\ref{11mar11:03}).
From (\ref{D1})\ we obtain $D_{1}\sim44.5$ so that $K_{\ast}=2$ and from
(\ref{15apr17:15}) we obtain $k_{\ast}=9.$ In a numerical simulation, we
started from the homogenous steady state $u=A,v=\frac{B}{A}$, perturbed by a
very small random noise. A Turing instability corresponding to the mode $k=7$
first develops. At time $t\sim50$ only six modes remain from which six mesas
develop. One by one, these mesas are annihilated until only two remain,
confirming our theory. We integrated the system until $t=\text{one million}$,
but we do not expect any more mesa refinement since $K_{\ast}=2$.

Formulas (\ref{15apr17:15}) together with (\ref{15apr17:20}) show that it is
possible for the mesa patterns to be stable even as the homogenous steady
state is stable. This occurs when $A^{2}/2<B<A^{2}$ (i.e. $l>\frac{1}{2}$)
with $\tau>\frac{B}{A^{2}}.$ A more difficult question is whether one can find
a regime in which both are \emph{unstable}, and the system iterates between
the two. Here we consider the case where $D$ satisfies (\ref{9mar14:03}). In
this regime the instability of a single mesa solution can only occur when
$\tau$ is near 1. Then (\ref{11mar11:03}) and (\ref{11mar11:04}) together
imply that $A\sim B^{2}.$ So we set:%
\begin{equation}
B=A^{2}+\alpha,\ \ \ \alpha\ll1.
\end{equation}
From the condition $T_{n=0}\leq0$ we obtain
\begin{equation}
\tau\geq1+\frac{\alpha-\varepsilon}{A^{2}}%
\end{equation}
and we suppose that a single mode $k=1\implies n=n_{\ast}=4\pi^{2}\varepsilon
D$ is unstable. The condition $\Delta_{n=n_{\ast}}<0$ then leads, to leading
order:%
\begin{equation}
4\pi^{2}\varepsilon D\left(  \varepsilon D4\pi^{2}-\alpha\right)  +\varepsilon
A^{2}<0
\end{equation}
so that to leading order,%
\begin{equation}
\alpha\geq\frac{A^{2}}{D4\pi^{2}}+4\pi^{2}\varepsilon D
\end{equation}
from where%
\begin{equation}
\tau\geq1+\frac{1}{D4\pi^{2}}\sim1+0.025\frac{1}{D}.
\end{equation}
On the other hand, we have $B\sim A^{2}\implies l=\frac{1}{\sqrt{2}};$ and we
have
\begin{equation}
ld-\frac{l^{3}+d^{3}}{3}=2l-2l^{2}-\frac{1}{3}=\sqrt{2}-1-\frac{1}{3}%
\end{equation}
so that from Theorem \ref{thm:hopf} we obtain
\begin{equation}
\tau_{h}\sim1+0.020\frac{1}{D}.
\end{equation}
Therefore the instability of a mesa cannot follow the Turing instability since
$\tau>\tau_{h}.$

\section{Discussion}

In Section \ref{sec:5} we were able to determine the instability thresholds
without actually computing the eigenvalues; but simply by showing the
existence of an asymmetric pattern bifurcating from a fold point. It is an
open problem to find the full expression for the eigenvalues near this
threshold. This would give a theoretical timescale for each of the step in the
coarsening process. We expect that the unstable eigenvalue will decrease
exponentially in the distance between the mesas. This would explain the
exponential time increase between the successive coarsening events, as
observed in Figure \ref{fig:turing}.a

In Theorem \ref{thm:hopf} under condition (\ref{9mar14:03}) we have shown that
as $\tau$ is decreased near 1, the first eigenvalue to cross the imaginary
axis is $\lambda_{+}$, whose eigenfunction is even. This corresponds to a
``breather''-type instability shown on Figure \ref{fig:turing}.b. An open
question is whether there exists a regime for which other eigenvalues undergo
a Hopf bifurcation before the $\lambda_{+}$ eigenvalue. For the Gray-Scott
model it is known that a single spike can undergo a Hopf bifurcation due to a
slow translational instability -- which corresponds to an odd eigenfunction --
whereby the center of the spike oscillates periodically \cite{Doelman},
\cite{KWW}, \cite{Muratov}. The analogy of this phenomenon for a single mesa
of the Brusselator would be the Hopf bifurcation of the $\lambda_{-}$
eigenvalue. One can also imagine spike-type solutions for the Brusselator
simply by taking the limit $l \rightarrow0$ or equivalently, $B \rightarrow
\infty.$ It is an open problem to study this regime.

It would be interesting to study the slow dynamics of the mesas, of which
there are several types, corresponding to different eigenvalues. The first
type is the slow translational motion of the mesa such as seen in Figure
\ref{fig:3} after time $t\sim2200$. Similar motion has been analysed for the
FitzHugh-Nagumo model on an infinite line \cite{GMP}. In contrast to the
Brusselator however, the FitzHugh-Nagumo model does not have a mass
conservation constraint $lK\sim\frac{A}{\sqrt{2B}}$ derived in Proposition
\ref{prop:1}, and does not undergo a coarsening process. In this sense the
Brusselator resembles more the Cahn-Hilliard model or the Allen-Cahn model
with mass constraint \cite{WardCH1}, \cite{WardCH2}. However unlike the
Cahn-Hilliard model, the Brusselator does not have a variational formulation,
and the mass conservation is only asymptotically valid. We remark that a
similar phenomenon was also studied for Gray Scott and Gierer-Meinhardt models
in the context of spike solutions \cite{IWW}, \cite{Doelman-speed},
\cite{Muratov-speed}.

A second type of slow instability is the mass exchange that occurs prior to
mesa annhiliation as seen in Figure \ref{fig:3} at time $t \sim2000$. This
phenomenon also occurs in some flame-propagation problems \cite{flame1},
\cite{flame2} and in the Keller-Segel model \cite{KKW}, where an exchange of
mass takes place between two boundary spikes, and eventually leads to an
annihilation of one of them.

The coarsening process in the brusselator terminates when there are
$K=O\left(  \frac{1}{\delta\ln\varepsilon^{-1}}\right)  $ mesas left, where
$\delta$ is the characteristic width of the interface. This is in contrast to
the the Cahn-Hilliard model, where the coarsening proceeds until all but one
interface remains \cite{WardCH1}.

Localized structures far from the Turing regime are commonplace in
reaction-diffusion systems such as the Brusselator, and provide an alternative
pattern-formation mechanism to Turing instability. These structures appear
whenever the Turing instability band is very large or when the diffusivity
ratio of the activator and inhibitor is large. As we demonstrate in this work,
Turing analysis cannot explain the diverse phenomena that can occur in this
regime, such as coarsening and the ``breather''-type instabilities. However
singular perturbation tools can be successfully applied to asnwer many of
these questions.

\section{Appendix A: proof of Lemma \ref{lemma:1}}

\textbf{Proof}. Note that (\ref{15dec1:49})\ is equivalent to solving%
\begin{align}
u^{\prime\prime}-\mu_{l}^{2}u  &  =0,\ \ \ x\in\cup\left(  x_{li}%
,x_{ri}\right) \\
u^{\prime\prime}-\mu_{d}^{2}u  &  =0,\ \ \ x\notin\cup\left(  x_{li}%
,x_{ri}\right) \\
u^{\prime}\left(  x_{li}^{+}\right)  -u^{\prime}\left(  x_{li}^{-}\right)   &
=-b_{li,}\ \ \ \ \ u^{\prime}\left(  x_{ri}^{+}\right)  -u^{\prime}\left(
x_{ri}^{-}\right)  =-b_{ri,}\\
u^{\prime}\left(  0\right)   &  =0=u^{\prime}\left(  1\right)  .
\end{align}

When $x\in\left[  x_{li,}x_{ir}\right]  $ we have%
\begin{equation}
u=u_{li}\cosh\left(  \mu_{2}\left(  x-x_{li}\right)  \right)  +B_{li}%
\sinh\left(  \mu_{2}\left(  x-x_{li}\right)  \right)  ,\ \ \ x\in\left[
x_{li,}x_{ir}\right]  \ \ \text{for}\ \ i=1\ldots K,
\end{equation}
where $u_{li}=u\left(  x_{li}\right)  $ and $B_{li}$ is to be found. We
similarly have%
\begin{equation}
u=u_{ir}\cosh\left(  \mu_{1}\left(  x-x_{ri}\right)  \right)  +B_{di}%
\sinh\left(  \mu_{1}\left(  x-x_{ri}\right)  \right)  ,\ \ \ x\in\left[
x_{ri,}x_{l\left(  i+1\right)  }\right]  \ \ \text{for}\ \ i=1\ldots K-1.
\end{equation}
We define%
\begin{equation}
d\equiv x_{r\left(  i+1\right)  }-x_{li}=\frac{1-l}{K},
\end{equation}%
\begin{align}
c_{1} &  \equiv\cosh\left(  \mu_{l}l\right)  ,\ \ s_{1}\equiv\cosh\left(
\mu_{l}l\right)  ,\\
c_{2} &  \equiv\cosh\left(  \mu_{d}d\right)  ,\ \ s_{2}\equiv\cosh\left(
\mu_{d}d\right)  .
\end{align}
We have $u_{ri}=u_{li}c_{l}+B_{li}s_{l},$ from where%
\begin{equation}
B_{li}=\frac{u_{ri}-u_{li}c_{l}}{s_{l}}~~\text{for}\ \ i=1\ldots K
\end{equation}
and similarly $u_{l\left(  i+1\right)  }=u_{ir}c_{d}+B_{ri}s_{d}$ so that%
\begin{equation}
B_{di}=\frac{u_{l\left(  i+1\right)  }-u_{ri}c_{d}}{s_{d}}~~\text{for}%
\ \ i=1\ldots K-1.
\end{equation}
We also have
\begin{align}
b_{li} &  =u^{\prime}\left(  x_{li}^{-}\right)  -u^{\prime}\left(  x_{li}%
^{+}\right)  =\mu_{d}\left(  u_{r\left(  i-1\right)  }s_{d}+B_{d\left(
i-1\right)  }c_{d}\right)  -\mu_{l}B_{li}\\
&  =\mu_{d}\left(  u_{r\left(  i-1\right)  }s_{d}+\frac{u_{li}-u_{r\left(
i-1\right)  }c_{d}}{s_{d}}c_{d}\right)  -\mu_{l}\frac{u_{ri}-u_{li}c_{l}%
}{s_{l}}\\
&  =-\frac{1}{s_{d}}\mu_{d}u_{r\left(  i-1\right)  }+\left(  \frac{c_{d}%
}{s_{d}}\mu_{d}+\frac{c_{l}}{s_{l}}\mu_{l}\right)  u_{li}-\frac{1}{s_{l}}%
\mu_{l}u_{ri}\ \ \text{for}\ \ i=2\ldots K
\end{align}
and similarly%
\begin{equation}
b_{ri}=-\frac{1}{s_{l}}\mu_{l}u_{li}+\left(  \frac{c_{d}}{s_{d}}\mu_{d}%
+\frac{c_{l}}{s_{l}}\mu_{l}\right)  u_{ri}-\frac{1}{s_{d}}\mu_{d}u_{r\left(
i+1\right)  }\ \ \text{for}\ \ i=1\ldots K-1.
\end{equation}
Next note that
\begin{equation}
u=A\cosh\left(  \mu x\right)  ,\text{ \ }x\in\lbrack0,x_{l1}]
\end{equation}
for some constant $A.$ Matching $u\left(  x_{l1}^{-}\right)  =u\left(
x_{l1}^{+}\right)  $ we then obtain
\begin{align}
u &  =\frac{u\left(  x_{l1}\right)  }{c_{d/2}}\cosh\left(  \mu x\right)  ,\\
u^{\prime}\left(  x_{l1}^{-}\right)   &  =u\left(  x_{l1}\right)  \mu
\frac{s_{d/2}}{c_{d/2}}.
\end{align}
where $s_{d/2}\equiv\sinh\left(  \mu_{d}d/2\right)  ,$\ $c_{d/2}\equiv
\cosh\left(  \mu_{d}d/2\right)  .$ \ Next we use the following identity,%
\begin{equation}
\frac{\sinh\left(  x/2\right)  }{\cosh(x/2)}=\frac{\cosh\left(  x\right)
-1}{\sinh(x)}=\frac{\sinh(x)}{1+\cosh(x)}%
\end{equation}
to obtain
\begin{equation}
u^{\prime}\left(  x_{l1}^{-}\right)  =\mu\frac{c_{d}-1}{s_{d}}u\left(
x_{l1}\right)  .
\end{equation}
Therefore we obtain%
\begin{align}
b_{l1} &  =\mu\frac{c_{d}-1}{s_{d}}u_{l1}-u^{\prime}\left(  x_{li}^{+}\right)
\\
&  =\mu\frac{c_{d}-1}{s_{d}}u_{l1}-\mu\frac{u_{r1}-u_{l1}c_{l}}{s_{l}}\\
&  =\mu\left(  \frac{-1}{s_{d}}\mu_{d}+\frac{c_{d}}{s_{d}}\mu_{d}+\frac{c_{l}%
}{s_{l}}\mu_{l}\right)  u_{l1}-\frac{1}{s_{l}}\mu u_{r1}%
\end{align}
and similarly%
\begin{equation}
b_{rK}=-\frac{1}{s_{l}}\mu_{l}u_{lK}+\left(  \frac{-1}{s_{l}}\mu_{l}%
+\frac{c_{d}}{s_{d}}\mu_{d}+\frac{c_{l}}{s_{l}}\mu_{l}\right)  u_{lK}.
\end{equation}
This yields the matrix $\mathbf{M}:$%
\begin{equation}
\mathbf{M}=\left[
\begin{array}
[c]{ccccccc}%
a+c & b &  &  &  &  & \\
b & c & a &  &  &  & \\
& a & c & b &  &  & \\
&  &  & \cdots &  &  & \\
&  &  &  & a & c & b\\
&  &  &  &  & b & c+a
\end{array}
\right]
\end{equation}
where
\begin{equation}
a=\frac{-\mu_{d}}{s_{d}},\ \ \ \ \ \ \ b=\frac{-\mu_{l}}{s_{l}}%
,\ \ \ \ \ \ \ c=\frac{c_{d}}{s_{d}}\mu_{d}+\frac{c_{l}}{s_{l}}\mu_{l}.
\end{equation}
Consider the matrix%
\begin{equation}
Q=\left[
\begin{array}
[c]{cccccccc}%
a & b &  &  &  &  &  & \\
b & 0 & a &  &  &  &  & \\
& a & 0 & b &  &  &  & \\
&  &  & \cdots &  &  &  & \\
&  &  &  & b & 0 & a & \\
&  &  &  &  & a & 0 & b\\
&  &  &  &  &  & b & a
\end{array}
\right]  .
\end{equation}
The eigenvalues of this matrix were computed in \cite{RW} (see Appendix B). It
was found that $Q$ has the following eigenvalues,%
\begin{gather}
\pm\sqrt{a^{2}+b^{2}+2ab\cos\left(  \theta\right)  },\ \ \theta=\frac{\pi
j}{K}\ \ \text{for}\ \ j=1\ldots K-1,\\
a+b,\ \ \ \ a-b.
\end{gather}
But w\bigskip e have $M=\left(  Q+c\right)  .\ $Therefore the eigenvalues of
$M$ are given by
\begin{gather}
\left(  c\pm\sqrt{a^{2}+b^{2}+2ab\cos\left(  \theta\right)  }\right)
,\ \ \theta=\frac{\pi j}{K}\ \ \text{for}\ \ j=1\ldots K-1,\\
c+a+b,\ \ c+a-b.
\end{gather}

\section*{Acknowledgments}
We would like to thank an anonymous referee and Y. Nishiura for suggestions that
helped to improve the manuscript.  This work was supported by the Fonds National de
la Recherche Scientifique (Belgium) and the InterUniversity Attraction Pole program
of the Belgian government.  The first author is grateful for the financial support
of NSERC PDF grant and Bourse de Post doctorat de l'ULB.

\end{document}